\documentclass[11pt]{article}                               

\usepackage{latexsym}
\usepackage{amssymb}
\usepackage{mathrsfs}
\usepackage{amsmath}
\usepackage{amsthm}
\usepackage{url}
\usepackage{color}
\usepackage{graphicx}
\usepackage{tikz}
\usepackage{jheppub}

\newcommand{\nc}{\newcommand}
\newcommand{\rnc}{\renewcommand}

\nc{\bea}{\begin{eqnarray}}
\nc{\eea}{\end{eqnarray}}
\nc{\be}{\bea}
\nc{\ee}{\eea}

\rnc{\a}{\alpha}
\nc{\ab}{\bar{\a}}
\nc{\ap}{\a^{+}}
\nc{\abm}{\ab^{-}}
\rnc{\b}{\beta}
\nc{\bb}{\bar{\b}}
\nc{\bbp}{\bb_{\zb}^{+}}
\nc{\bm}{\b_{z}^{-}}
\nc{\oa}{\overline{\a}}
\nc{\ob}{\overline{\b}}
\rnc{\gg}{\gamma}
\rnc{\d}{\delta}
\nc{\f}{\phi}
\nc{\fb}{\bar{\phi}}
\nc{\vf}{\varphi}
\nc{\p}{\psi}

\rnc{\c}{\chi}
\nc{\la}{\lambda}
\nc{\m}{\mu}
\nc{\n}{\nu}
\nc{\Om}{\Omega}
\rnc{\t}{\theta}
\nc{\eps}{\epsilon}
\rnc{\S}{\Sigma}
\nc{\F}{\Phi}

\nc{\trac}[2]{{\textstyle\frac{#1}{#2}}}

\nc{\ex}[1]{\mbox{e}^{\,\textstyle#1}}

\nc{\mat}[4]{\left(\begin{array}{cc}#1&#2\\#3&#4\end{array}\right)}

\nc{\som}[9]{\left(\begin{array}{ccc}#1&#2&#3\\#4&#5&#6\\#7&#8&#9%
\end{array}\right)}

\nc{\ad}{\mathop{\mathrm{ad}}\nolimits}
\nc{\Ad}{\mathop{\mathrm{Ad}}\nolimits}
\nc{\tr}{\mathop{\mathrm{tr}}\nolimits}
\nc{\Tr}{\mathop{\mathrm{Tr}}\nolimits}
\nc{\Det}{\mathop{\mathrm{Det}}\nolimits}
\nc{\rk}{\mathop{\mathrm{rk}}\nolimits}
\nc{\ra}{\rightarrow}
\nc{\Ra}{\Rightarrow}
\nc{\LRa}{\Leftrightarrow}
\nc{\ot}{\otimes}
\rnc{\ss}{\subset}
\nc{\nul}{\noindent\underline}
\nc{\non}{\nonumber\\}

\nc{\zb}{\bar{z}}
\rnc{\lg}{\mathfrak{g}}
\nc{\lt}{\mathfrak{t}}
\nc{\lk}{\mathfrak{k}}
\nc{\lig}{\mathfrak{ig}}
\nc{\lh}{\mathfrak{h}}

\nc{\CC}{\mathbb{C}}
\nc{\EE}{\mathbb{E}}
\nc{\RR}{\mathbb{R}}
\nc{\ZZ}{\mathbb{Z}}
\nc{\NN}{\mathbb{N}}

\newtheorem{theorem}{Theorem}[section]

\newtheorem{fact}{Fact}

\newtheorem{example}[theorem]{Example}

\makeatletter
\gdef\@fpheader{}
\makeatother

\title{On the Evaluation of the Ray-Singer Torsion Path Integral}
\author[a]{Matthias Blau}
\author[b,c]{Mbambu Kakona}
\author[c]{George Thompson}

\affiliation[a]{Albert Einstein Center for Fundamental Physics, Institute of
Theoretical Physics, University of  Bern, Sidlerstrasse 5, CH-3012
Bern, Switzerland.}

\affiliation[b]{
East African Institute for Fundamental Research (EAIFR), 
University of Rwanda, KN 7 Ave,
Kigali, 
Rwanda.}

\affiliation[c]{
Abdus Salam International Centre for Theoretical Physics, 
Strada Costiera 11, 
34014 Trieste, 
Italy.}

\emailAdd{blau@itp.unibe.ch}
\emailAdd{kakona@eaifr.org}
\emailAdd{thompson@ictp.it}

\abstract{There are very few explicit evaluations of path integrals for 
topological gauge theories in more than 3 dimensions. Here we  provide
such a calculation for the path integral representation of the Ray-Singer
Torsion of a flat connection on a vector bundle on base manifolds that are 
themselves $S^{1}$ bundles of any dimension. The calculation relies on
a suitable algebraic choice of gauge which leads to a convenient factorisation of
the path integral into horizontal and vertical parts.

}

\global\parskip=6pt

\begin{document}
\maketitle

\setcounter{footnote}{0}

\section{Introduction}

There are very few explicit evaluations of path integrals for 
topological gauge theories in more than 3 dimensions. 
Here we 
will provide such calculations for a class of topological field
theories giving a path integral realisation of the Ray-Singer Torsion.
The Ray-Singer Torsion \cite{Ray-Singer, Ray-Singer-Analytic} is an
invariant that can be defined on any compact closed manifold $M$ of
real dimension $\dim_{\mathbb{R}}M=m$. The
definition requires a bundle $E$ over $M$ that admits a flat
connection. Schwarz \cite{Schwarz-Partition, Schwarz,Schwarz-Tyupkin} gave a deceptively easy
topological field theory 
representation for the Torsion. The Schwarz action for the Ray-Singer
Torsion of a flat connection $\mathcal{A}$ on a bundle $E\rightarrow M$
is given by
\be
S_\mathcal{A}= \int_{M} \Tr B\wedge \, d_{\mathcal{A} }\,
C \label{schwarz}
\ee
with $
 B \in \Omega^{p}(M, E^{*})$, $C \in
 \Omega^{q}(M, E)$
where $q=m-p-1$. The action (\ref{schwarz}) has reducible symmetries
  as $d_{\mathcal{A} }^{2}=F_\mathcal{A}=0$. Schwarz formally
  quantised these reducible systems in terms of
resolvents. The resolvent approach was
shown to be equivalent to the more usual Faddeev-Popov gauge fixing
procedure in \cite{BT-RST,BT-BF, BT-Metric}. Either way the partition function $Z[\mathcal{A},E]$ with
action (\ref{schwarz}) is related to the Ray-Singer torsion
$\tau_{M}(\mathcal{A},E)$ by
\be
Z[\mathcal{A}, E] =
\tau_{M}(\mathcal{A},E)^{(-1)^{p+1}} \label{part-tor}
\ee

In \cite{BKT} we began a study of the exact evaluation of the Schwarz
path integral representation of the Ray-Singer Torsion. Particular
emphasis was paid to the issue of zero modes and we advocated using
the addition of a mass to take these into account. Formally, the
addition of a mass is straightforward, however, computationally we
focused on mapping tori as on these manifolds the mass could be understood as a
part of a connection for an extended gauge group. Using path integral techniques
we were able to reproduce for the Ray-Singer Torsion a result of Fried
\cite{Fried} that was established for the Reidemeister Torsion. That
Fried's result should also hold for Ray-Singer Torsion is a
consequence of the equivalence of the two Torsions conjectured by Ray
and Singer and proven by Cheeger
\cite{Cheeger-1,Cheeger-2} and by M\"{u}ller \cite{Muller}.

Here we consider the same path integrals but for the manifold $M$ taken to be an $S^{1}$ bundle
over $N$ and  where $E$ is a flat $G=SU(r)$ vector bundle over $M$. We
provide an explicit evaluation of the path integrals involved for $M$
of any dimension and $E$ of any rank.

The
natural nowhere vanishing vector field on $M$ generating the $S^{1}$-action allows for a
decomposition of any connection $\mathcal{A}$ on $E$ into a component $\phi$ 
along the fibre and a horizontal component $A$. By the same
logic the fields $B$ and $C$ and their entire Batalin-Vilkovisky
triangles of ghost and multiplier fields \cite{BV1983} may also be decomposed into
components. This allows one to impose an algebraic gauge condition
which both simplifies the field content and simultaneously allows one
to perform the path integrals. 

We find that quite generally
(\ref{part-tor}) factorises as
\be
Z[\mathcal{A}, E] = Z_{\parallel}[A, E] \, . \, Z_{\perp}[\phi, E] \label{intro-factor}
\ee
where $Z_{\parallel}[A, E] $ contains all the information about
the zero modes associated with the flat connection. Hence in the acyclic
case (and up to trivial normalisation) $Z_{\parallel}[A, E] =1$,
otherwise $Z_{\parallel}[A, E] $ is a section of an appropriate
determinant line bundle. Whether a flat connection which is acyclic
exists or not depends on the $SU(r)$ bundle $E$. For example if $E$ is the fundamental
representation then there are flat connections which are acyclic whereas
for the adjoint bundle case no flat connection is acyclic. As how to deal with zero
modes has been explained previously \cite{BT-BF, BKT} we do not spend much time on this
part of the factorisation formula.

The more interesting factor is $Z_{\perp}[\phi, E]$ and to evaluate it
we make some choices. We presume that $E$ is the pullback of a bundle
$V$ on $N$. On $N$, $A$ is a connection while $\phi$ is a section of
the $\ad V$ bundle. In this case the flatness equations for the combined gauge field
on $E$ over $M$ imply
that either $\phi$ is zero (in which case $A$ is a flat connection on
$N$) or $\phi$ is non-zero and $A$ is a reducible connection. If $\phi$
is zero there is nothing to do, so we fix our attention on the case
where $\phi$ is not zero. In this case the flatness
condition implies that the connection $A$
splits as $A^{K} \oplus A^{U(1)}$ with $A^{U(1)}$ lying in the same
direction as $\phi$ in the Lie algebra and $K \times U(1) \subset
G$. With this set-up the computations are quite doable and one finds
that
\be
Z_{\perp}[\phi, E]=\left(\tau_{S^{1}}(\phi,
  E_{\perp}|_{S^{1}})^{\chi(N)}\right)^{(-1)^{p+1}} \label{pert-top}
\ee
where $E_{\perp}|_{S^{1}}$ is the restriction to the fibre of the
bundle $E_{\perp}$ over $M$, $\tau_{S^{1}}$ is the Ray-Singer
torsion on the circle and $\chi(N)$ is the Euler characteristic of
$N$. (The definition of $E_{\perp}$ is given in 
section \ref{section6}; here one understands that in (\ref{pert-top})
it corresponds to having no zero modes.)

Combining (\ref{part-tor}) and (\ref{pert-top}) in the acyclic case
we arrive at
\be
\tau_{M}(\mathcal{A},E) = \tau_{S^{1}}(\phi,
  E_{\perp}|_{S^{1}})^{\chi(N)}
\label{introfried}
\ee
which agrees with another result of Fried for Reidemeister Torsion (result
(V) on page 26 of \cite{Fried}) when $M$ is an $S^{1}$ fibration. Our results
for the non-acyclic case therefore represent a generalisation of that
work. 

While this may be of some limited interest in its own right, the ability
to reproduce or generalise known mathematical results serves mainly
as a check on the methods that we develop and use here to evaluate the path integral, 
which can then also be applied with more confidence in other cases.

In Section \ref{section2}, in order to set the stage, we review the
flatness equations for connections on a vector bundle over an
$S^{1}$ bundle, while a more general background on the geometry of
the bundles and the decomposition that we use are provided in
Appendix \ref{appendix1}. In particular, we explain how to decompose
forms, exterior (covariant) derivatives, connections and curvatures into
horizontal and vertical components.

Section \ref{section3} is devoted to establishing that the
partition function factorises as in (\ref{intro-factor}). In the current geometric situation
it is possible to impose algebraic gauge fixing conditions that
simplify the ghost for ghost system of the quantisation of the theory
and we carry this out in detail in
Appendix \ref{appendix2}.  With the gauge fixing choices that we make the 
factorised partition functions have the dependency on the components of
the connection explicitly exhibited in
(\ref{intro-factor}). 

An analysis of the factor $Z_{\parallel}[A, E] $ is the subject of
Section \ref{section4}. There it is shown that the path integrals
devolve to integrals over the zero modes or, equivalently, to
producing a section of a determinant line bundle of tensor powers of cohomology
groups associated with the twisted de Rham operator. Such determinant
lines appear in the work of Ray and Singer \cite{Ray-Singer-Analytic} and from the quantum
gauge theory point of view \cite{BT-BF, BKT} so we are brief about
this here. However, we do spend some time on establishing which harmonic modes
are actually zero modes and which are part of the gauge symmetry.

In Section \ref{section5} we turn our attention to the second
factor and begin a first evaluation of $Z_{\perp}[\phi, E] $ in
generality, that is before making specific choices of the bundles and
connections on them. We show that $Z_{\perp}[A, E] $
is essentially products of powers of the determinant of the covariant
derivative in the fibre direction $D_{\phi}$ acting on the spaces
spanned by the forms $B$, $C$ and their ghost systems. At this general level one cannot go too far but
we do obtain a formal expression for the partition function that
suggests that the final answer only depends on cohomology
groups with values in certain vector bundles over $N$.

In order to proceed further with the analysis of the partition
function $Z_{\perp}[\phi, E] $ we take a more in-depth look at the
flatness equations in Section \ref{section6}. The flat bundle $E$ 
over $M$ that is considered is a pullback of a bundle $V$ over $N$,
so that $E = \pi^{*}V$. In this situation $A$ is the pullback of a
connection on $V$ that is not flat, $\phi$ is the pullback of a section
of the $\ad V$ bundle while the flat connection on $E$ is a
combination of these two. We also assume that $\phi$ is not zero so
that, as described above, the structure group 
reduces to $K \times U(1) \subset G$ and that the connection splits as
$A^{K} \oplus A^{U(1)}$. In such a situation $V$ splits as the direct
sum of powers of a basic line bundle $\mathcal{L}$ over $N$ and the
curvature 2-form of the connection $A^{K}$ is flat on $N$.

Section \ref{section7} is where we put all of the pieces together and
determine $Z_{\perp}[\phi, E] $ for the choices of the previous
section. The determinants of $D_{\phi}$  found in Section \ref{section5} are now over
spaces of forms on $N$ with values in tensor powers of the line bundle
$\mathcal{L}$. The combinations for fixed Fourier mode (on the fibre)
are regularised as a Heat Kernel and the combination is the super-trace
of the eigenmodes of the twisted de Rham operator, which in turn is a
multiple of the Euler characteristic $\chi(N)$ of $N$, which explains
its appearance in (\ref{pert-top}). The product over the Fourier modes
is regularised using a $\zeta$ function regularisation and provides
the $\tau_{S^{1}}$ that appears in (\ref{pert-top}).

Finally, in Section \ref{section8} we briefly mention certain
generalisations that can be considered, some of which straightforwardly
follow from the computations that we have presented here.

There are two appendices, as mentioned above. In the  first Appendix
\ref{appendix1} we provide a quick summary of the geometric structure
that is available on an $S^{1}$ bundle $M\to N$. This includes
the decomposition of forms and connections into horizontal and
vertical components. Amongst other things 
we also review 
the relationship with basic cohomology 
and explain how to expand fields on $M$ in ``Fourier modes" as sections of
line bundles over $N$. 

The algebraic gauge
conditions that we will make use of and that simplify our
considerations in the body of the paper are introduced in Appendix \ref{appendix2}. We explain 
how such algebraic choices allow one to avoid the complete set of ghosts for ghosts structure that
would normally be required for the reducible symmetries of the theory
in covariant gauges.

\section{Connections on Bundles over \texorpdfstring{$\boldsymbol{S^{1}}$}{S1} Bundles}\label{section2}

Let $M$ be the total space of a smooth $S^1$-bundle over a smooth manifold $N$. 
We will need to make use of some aspects of the geometry of $S^1$-bundles. 
The results that we need are quite standard, so we do not repeat them here, but
for convenience we have assembled and briefly reviewed them in Appendix \ref{appendix1}.

Thus let $\pi: M \to N$ be a smooth $S^1$-bundle over the smooth manifold $N$, 
and denote by $\xi$ the fundamental nowhere vanishing vector field generating the 
$S^1$-action on $M$. A choice of connection on this bundle is equivalent to a
choice of 1-form $\kappa$ on $M$ satisfying 
\be
\iota_\xi \kappa = 1 \quad,\quad \mathcal{L}_\xi \kappa = 0 \quad\Ra\quad \iota_\xi d\kappa = 0
\label{kappa1}
\ee
(with $\mathcal{L}_{\xi} = \iota_{\xi} d + d \iota_{\xi}$ the Lie derivative), 
and we choose one such $\kappa$ in the following. 

With respect to a choice of $\kappa$, any 
$p$-form $\alpha\in \Omega^{p}(M)$ on $M$ can be decomposed into a
``horizontal'' and a ``vertical'' component,  
\be
\alpha = (\iota_\xi \kappa) \alpha  = \iota_\xi(\kappa\wedge \alpha) + \kappa
\wedge \iota_\xi\alpha \equiv 
\alpha^{(p)} + \kappa \wedge \alpha_{(p-1)} 
\label{formsplit}
\ee
with $\alpha^{(p)} = \iota_\xi (\kappa\wedge\alpha)$ and
$\alpha_{(p-1)} =\iota_\xi\alpha$ 
both manifestly horizontal, in the sense that 
\be
\iota_{\xi}\alpha^{(p)}=0\quad, \quad \iota_{\xi} \alpha_{(p-1)}=0\;\;.
\ee
The space of horizontal $p$-forms $\alpha$, i.e.\  which satisfy
$\iota_{\xi}\alpha =0$, is denoted by $\Omega_{H}^{p}(M)$. 
Likewise, the exterior derivative can be decomposed as \eqref{split-deriv}
\be
d = \iota_\xi (\kappa \wedge d) + \kappa \wedge \iota_\xi d \equiv d_H + \kappa
D\;\;.
\label{2dsplit}
\ee

Now let $E$ be a vector bundle over $M$. A connection
$\mathcal{A}$ on $E\rightarrow M$ may be decomposed, as above, as
\be
\mathcal{A} = A + \kappa \phi, \quad \iota_{\xi}A =0 
\ee
so that locally $A$ is a Lie algebra valued horizontal 1-form and
$\phi$ is, locally, a Lie algebra valued 0-form. Gauge transformations
decompose appropriately,
\be
A \rightarrow g^{-1}Ag + g^{-1}d_{H}g,  \quad \phi  \rightarrow
g^{-1}\phi g + g^{-1}\mathcal{L}_{\xi}g  
\ee
and covariant derivatives split as in \eqref{2dsplit}, i.e.\ one has
\be
d_{\mathcal{A}} = d + \mathcal{A} \; = d_{A}^{H} +
\kappa D_{\phi}, \quad D_{\phi}= D + \phi, 
\label{dasplit}
\ee
(with the connection in the appropriate representation), and 
the square of the covariant derivative is 
\begin{align}
d_{\mathcal{A}}^{2} & = \left( d_{A}^{H}\right)^{2} + \left\{ d_{A}^{H} ,
  \kappa D_{\phi}\right\}\notag\\
&= \left( d_{A}^{H}\right)^{2}  + d\kappa D_{\phi} - \kappa
[d_{A}^{H} ,D_{\phi}] \label{split-deriv-sq}
\end{align}
The corresponding decomposition of the curvature 2-form is
\be
F_{\mathcal{A}} = F_{A}^{H} + d\kappa \phi + \kappa
\left(\mathcal{L}_{\xi}A - d_{A}^{H} \phi
\right) \label{curvature}
\ee
where for simplicity we have adopted the notation
\be
F_{A}^{H} = d_{H}A +A^{2}, \quad
d_{A}^{H} \phi = d_{H}\phi + [A, \phi]
\ee
Note that $F_{A}^{H}$ does not satisfy the `horizontal' Bianchi
identity $d_{A}^{H}F_{A}^{H}=0$ in general as $d_{H}^{2}\neq 0$. However,
if $\mathcal{L}_\xi A=0$, one has 
\be
\mathcal{L}_{\xi}A=0 \;\; \Rightarrow \;\; d_{H}A =dA \quad \Rightarrow 
\quad F_{A}^{H}=F_{A} 
\ee
and thus this condition implies the horizontal Bianchi identity, 
\be
\mathcal{L}_{\xi}A=0 \;\; \Rightarrow \;\; 
d_A^H F_A^H = d_AF_A =0\;\;.
\label{Bianchi}
\ee
If this condition is satisfied then one can consider $A$ to be the pullback of a
connection (which we also denote by $A$) on a bundle on $N$. The pullback
curvature $F_{A}^{H}$ now correctly satisfies the appropriate Bianchi
identity.

The equations for a flat connection are, from (\ref{curvature}),
\be
F_{A}^{H} =-  d\kappa \, \phi , \quad \mathcal{L}_{\xi}A = d_{A}^{H} \phi \label{flat-1}
\ee
The `operator' version of these equations that follow from (\ref{split-deriv-sq})
are
\be
\left(d_{A}^{H}\right)^{2}  =- d\kappa D_{\phi} , \quad \kappa \wedge
[d_{A}^{H} ,D_{\phi}] =0 \label{split-deriv-sq-2}
\ee
and will be useful in the next section. In particular, the second 
equation implies that the operators $d_A^H$ and $D_\phi$ commute 
on horizontal vector-valued forms $\alpha \in \Omega_{H}^{p}(M,E)$, 
\be
\alpha \in \Omega_{H}^{p}(M,E) \quad\Rightarrow\quad 
[d_{A}^{H} ,D_{\phi}]\,\alpha = 0 \label{commute}
\ee

\section{Factorisation of the Partition Function}\label{section3}

Let $\mathcal{A}= A + \kappa \phi$ be a flat connection on a vector bundle
$E$. The Schwarz action for the corresponding Ray-Singer torsion is the 
action (\ref{schwarz}),
\be
S_{\mathcal{A}}  = \int_{M}\langle B, \,  d_{\mathcal{A}}C\rangle
\label{eschwarz}
\ee
with the fields in $\Omega^*(M,E)$. Here we use $\langle \cdot , \cdot
\rangle$ to denote a fixed 
invariant fibre scalar product on $E$ (which in (\ref{schwarz}) was
denoted by $\Tr$). In case $E$ is a complex vector
bundle it is to be understood throughout that we have added the
complex conjugates of the fields in (\ref{eschwarz}) so that the
action is real. Specific choices 
are e.g.\ the bundle $E=F$ for the fundamental representation of some group $G$, or 
$E = \ad P$ for the adjoint representation (associated to the principal $G$-bundle
$P$). These are the examples that we will occasionally consider explicitly.
However, we can describe both the gauge fixing and the evaluation of the 
path integral in a way that is largely independent of the choice of $E$ (or
representation).  

The gauge fixing is described in detail in Appendix  \ref{appendix2}
for any $E$ equipped with a flat connection.  The gauge choices are
a mixture of algebraic conditions (imposed directly on components
of the fields and do not involve differential operators) and those
that do involve differential operators (as in the Landau gauge in
Yang-Mills theory, say).

There is a good reason to use such algebraic gauges in the present context.  These
gauge conditions greatly simplify the evaluation of the path integral
and because of that simplification lead directly to the somewhat
startling fact that the partition function factorises into two pieces,
one piece only contains possible zero mode dependence and otherwise is 
independent of the $A$ part of the background flat connection $\mathcal{A}=
A + \kappa \phi$. The second factor, one sees in a very direct
fashion,  depends
only on the $\phi$ component of the connection. Denoting the complete 
partition function by $Z[\mathcal{A},E]$, one thus has the factorisation
\be
Z[\mathcal{A},E]= Z_{\parallel}[A]\, . \, Z_{\perp}[\phi]\label{Z-split}
\ee
(the notation will become apparent). To exhibit this simple
dependence on $\phi$ one does not  
need to know the details of the vector bundle nor the details
of the flat connection under consideration.

The key step in order to establish this factorisation (and, as it
turns out, to neatly separate zero modes from non-zero modes) is
to perform a decomposition of horizontal forms according to the
action of $D_\phi$ on them.
To that end, let
\be
\Omega^{p}_{H}(M,E) = \Omega^{p}_{\parallel}(M,E) \oplus
\Omega^{p}_{\perp}(M,E)\label{par-perp} 
\ee
where $\Omega^{p}_{\parallel}$ is the kernel of the map
\be
D_{\phi}:\Omega^{p}_{H}(M,E) \rightarrow\Omega^{p}_{H}(M,E)
\ee
and $D_{\phi}$ is invertible on $\Omega^{p}_{\perp}(M,E)$. By virtue
of (\ref{commute}) one also has that
\be
d_{A}^{H}:\Omega^{p}_{\parallel}(M,E) \rightarrow \Omega^{p}_{\parallel}(M,E)
\ee
For a horizontal form $\alpha \in \Omega^{p}_{H}(M,E)$ we write
\be
\alpha = \alpha_{\parallel} + \alpha_{\perp}
\ee
with the obvious notation. As in \eqref{formsplit}, the fields
and the ghosts etc have the decomposition 
\be
\Psi\in \Omega^{q}(M,E) \quad\Rightarrow\quad 
\Psi = \Psi^{(q)} + \kappa \Psi_{(q-1)}
\ee
with $\Psi^{(q)}$ and $\Psi_{(q-1)}$ horizontal. 
As one sees from (\ref{red-symm}) and (\ref{red-symm-2}), or in more
detail in (\ref{symm2}), the field
$\Psi_{(q-1)}$ will transform as 
\be
\Psi_{(q-1)}\rightarrow \Psi_{(q-1)} + D_{\phi} \Lambda_{\perp}^{(q-1)}
\ee
where we have chosen the gauge parameter $\Lambda_{\perp}^{(q-1)} \in
\Omega^{q-1}_{\perp}(M,E)$. Taking $\Lambda_{\perp}^{(q-1)}= -
D_{\phi}^{-1}\Psi^{\perp}_{(q-1)}$ allows one to remove 
the $\Psi^\perp_{(q-1)}$ from $\Psi_{(q-1)}$, so that only
the parallel component survives, 
\be
\Psi_{(q-1)}\rightarrow \Psi_{(q-1)}^{\parallel}
\ee
Now we can state the algebraic gauge condition: for all fields $\Psi$
we impose the condition that
\be
\Psi_{(p-1)}^{\perp}=0 \label{perp-gf}
\ee
In particular, for the fields
  \be
B = B^{(p)}+
\kappa B_{(p-1)} , \quad C = C^{(q)}+ \kappa C_{(q-1)}
\ee
in the action (\ref{eschwarz}), the algebraic conditions are
\be
B_{(p-1)} ^{\perp}=0, \quad C_{(q-1)}^{\perp}=0
\ee
and
the classical action \eqref{eschwarz} for the vector bundle $E$
then becomes
\begin{align}
S_{\mathcal{A}} &= \int_{M} \left( \langle B^{(p)}_{\perp}, \, \kappa D_{\phi} C^{(q)}_{\perp}\rangle
  -\langle B^{(p)}_{\parallel}, \, \kappa
                  d^{H}_{A}C_{(q-1)}^{\parallel} \rangle \right.\nonumber
  \\
  & \left. \quad  \quad + 
     \kappa
        \langle B_{(p-1)}^{\parallel} , \, d^{H}_{A}
    C^{(q)}_{\parallel} \rangle + \kappa d\kappa
        \langle B_{(p-1)}^{\parallel}, \, C_{(q-1)}
    ^{\parallel}\rangle \right) \label{action-split}
\end{align}
with the fields being sections of the appropriate bundles. Notice that
`parallel' and `perpendicular' fields do not couple. 

There are still more symmetries to gauge fix. We have only used up the
$\Omega^{*}_{\perp}(M,E)$ parts of the symmetry. If $\Psi$ satisfies
(\ref{perp-gf}) then the transformation
\be
\Psi^{(p)}_{\parallel} \rightarrow \Psi^{(p)}_{\parallel}  +
d_{A}^{H}\Lambda^{(p-1)}_{\parallel} + d\kappa
\Lambda_{(p-2)}^{\parallel} , \quad  \Psi_{(p-1)}^{\parallel} \rightarrow \Psi_{(p-1)}^{\parallel} -
d_{A}^{H} \Lambda_{(p-2)}^{\parallel}  \label{parallel-trans}
\ee
will preserve the gauge condition. These symmetries only involve the
fields in $\Omega^{*}_{\parallel}(M,E)$ and so gauge fixing these
symmetries does not mix with the gauge fixing of the perpendicular
symmetries. Consequently we conclude that for the complete partition
function the total action including ghosts and other fields will split
precisely into perpendicular and parallel components just as in
(\ref{action-split}) and this is borne out in some detail in Appendix
\ref{appendix2}. This gauge fixing and ghost structure therefore guarantees that (\ref{Z-split}) holds.

\section{\texorpdfstring{$\boldsymbol{Z_{\parallel}[A]}$}{Z[A]} and the Zero Mode Measure}\label{section4}

One can be somewhat more precise about the structure of
$Z_{\parallel}[A]$ and at the same time come to a better
understanding of the $d\kappa$ component of the gauge transformations.

We gauge fix in two steps. Firstly note that by (\ref{parallel-trans}) we may
covariantly gauge fix both the `upstairs' parallel fields
$\Psi^{(p)}_{\parallel} $ and the `downstairs' $\Psi_{(p-1)}^{\parallel} $ ones as
\be
d_{A}^{H} *_{H}\Psi^{(p)}_{\parallel} =0, \quad d_{A}^{H}
*_{H}\Psi_{(p-1)}^{\parallel} =0 \label{cov-gf}
\ee
There is still a residual symmetry that is available even after we
have used $\Lambda_{(p-2)}^{\parallel} $ in order to impose the second
condition in (\ref{cov-gf}). Those $\Lambda_{(p-2)}^{\parallel} $
which are harmonic with respect to $\Delta_{A}^{H}$ (so those that
satisfy $d_{A}^{H} \Lambda_{(p-2)}^{\parallel} =0$ in the
transformation (\ref{parallel-trans}) as well as $d_{A}^{H}
*_{H}\Lambda_{(p-2)}^{\parallel} =0$ so as to preserve the condition
on $\Psi_{(p-1)}^{\parallel} $ in  (\ref{cov-gf})) are still
available.

Likewise there are $\Lambda^{(p)}_{\parallel} $
harmonic modes. We will discuss the roles of the harmonic modes and
their relationship with zero modes below.

All fields in $\Omega_{\parallel}^{*}(M,E)$ are either $d_{A}^{H}$
exact or $\d^{H}_{A}$ exact or harmonic with respect to
$\Delta_{A}^{H}$. This follows from the fact that $(d_{A}^{H})^{2}=0
$ on $\Omega_{\parallel}^{*}(M,E)$
as follows from  
\eqref{split-deriv-sq-2}, 
\be
\left(d_{A}^{H}\right)^{2}\alpha_{\parallel}  =- d\kappa
D_{\phi}\alpha_{\parallel} = 0 
\ee
and that with the respect to the
inner product (\ref{inner-hodge}) this Hodge decomposition is an
orthogonal one. With this in mind it is convenient  to set
\be
Z_{\parallel}[A]= \mu \, . \, \widehat{Z}_{\parallel}[A] \label{zero-mode-mu}
\ee
where $\mu $ is a section of an appropriate determinant line bundle of
cohomology groups and
$\widehat{Z}_{\parallel}[A]$ is the path integral we had before except
with the harmonic mode sector projected away.  $\mu$ contains all the
information of the harmonic modes and is there to ensure that the
overall partition function is metric independent. It is important to
note that $Z_{\perp}[\phi]$ on the other hand carries no 
zero modes as it only involves the perpendicular fields.

We now establish that $ \widehat{Z}_{\parallel}[A]=1$. In
Appendix \ref{appendix2} we show that in the gauge transformations
(\ref{parallel-trans}), in the sector orthogonal to harmonic modes,
the $d\kappa$ terms can be ignored.  We give a
slightly different argument to this effect here. Start with the
action (\ref{action-split}), with all fields orthogonal to the
harmonic modes, and send
\be
B^{(p)}_{\parallel} \rightarrow t B^{(p)}_{\parallel} , \quad
C^{\parallel}_{(q-1)} \rightarrow t^{-1} C^{\parallel}_{(q-1)} 
\ee
so that the last term goes over to
\be
t^{-1}\int_{M} \kappa \, d\kappa \, \langle B^{\parallel}_{(p-1)}, \, 
C^{\parallel}_{(q-1)}\rangle 
\ee
while the others do not change. Now on taking the $t\rightarrow
\infty$ limit this last term completely vanishes. For consistency,
however, we also need to scale all the fields in the symmetry
transformations. The overall effect is to send the $d\kappa
\Lambda_{(p-2)}^{\parallel} $ in (\ref{parallel-trans}) to $t^{-1}d\kappa
\Lambda_{(p-2)}^{\parallel} $ so that this too vanishes in the
limit. One must bear in mind that these scalings are done orthogonally
to the harmonic modes, so that each mode under discussion here appears
in the action.

At this point the action for the partition function $
\widehat{Z}_{\parallel}[A]$ is
\be
S_{A}^{\parallel}= \int_{M}\left(
  -\langle B^{(p)}_{\parallel}, \, \kappa
  d^{H}_{A}C_{(q-1)}^{\parallel}\rangle + 
     \kappa
        \langle B_{(p-1)}^{\parallel}, \,  d^{H}_{A} C^{(q)}_{\parallel}\rangle
      \right) \label{action-parallel}
\ee
while the symmetries have gone over to
\be
\Psi^{(p)}_{\parallel} \rightarrow \Psi^{(p)}_{\parallel}  +
d_{A}^{H}\Lambda^{(p-1)}_{\parallel} , \quad  \Psi_{(p-1)}^{\parallel}
\rightarrow \Psi_{(p-1)}^{\parallel} - 
d_{A}^{H} \Lambda_{(p-2)}^{\parallel}  \label{parallel-trans-2}
\ee      
At the outset, the $d\kappa$ part of the symmetries were there as
$(d_{A}^{H})^{2} \neq 0$. One may therefore wonder how we have managed to 
get away with eliminating those pieces. Essentially, this is due to the fact
that on parallel fields, as we have already seen, $(d_{A}^{H})^{2}=0$
so that on those fields $d_{A}^{H}$ is essentially flat. In this way we understand that
(\ref{action-parallel}), which only contains parallel fields,
and the symmetries (\ref{parallel-trans-2}) (which exclude the
harmonic mode symmetries)
correspond to two Schwarz actions of Ray-Singer type for the `flat'
connection $d_{A}^{H}$ . We thus see that the partition function $\widehat{Z}_{\parallel}[A]$
itself factorises into two parts, one for the $(B^{(p)}_{\parallel},
C_{(q-1)}^{\parallel})$ system and the other for the $(B_{(p-1)}^{\parallel} ,
C^{(q)}_{\parallel})$ system.

The path integrals in question need regularisation and one
may adopt the original definition in terms of $\zeta$-functions of Ray
and Singer \cite{Ray-Singer} where the Laplacian in question is
$\Delta_{A}^{H}=d_{A}^{H}\delta_{A}^{H}+ \delta_{A}^{H}d_{A}^{H}$ where
$\delta_{A}^{H}$ is the $*_{H}$ dual of $d_{A}^{H}$. Now with this
definition a general $(B^{(p)}_{\parallel},
C_{(q-1)}^{\parallel})$ system yields $\tau(A)^{(-1)^{p-1}}$ where
$\tau(A)$ is the `Ray-Singer torsion' of the connection $A$. Consequently, the product of
these two partition functions cancel and we are left with unity. Thus
\be
\widehat{Z}_{\parallel}[A]=1
\ee
as we set out to show. In particular the definition used by Ray and
Singer in \cite{Ray-Singer} was to project out the harmonic modes just
as we have done here.

Clearly if all of the
$\Omega^{*}_{\parallel}(M,E)$ are trivial then (up to normalisation)
one may set $Z_{\parallel}[A]=1$. This fact is related to the
acyclicity of the de Rham complex $(d_{\mathcal{A}}, E)$. The twisted de Rham
complex is said to be acyclic if all the twisted de
Rham groups $\mathrm{H}_{\mathcal{A}}^{*}(M,E)$ are trivial. The
representatives of the twisted de Rham groups satisfy
\be
d_{\mathcal{A}}v=0 , \quad \Rightarrow \quad d_{A}^{H}v^{(p)} + d\kappa\wedge
v_{(p-1)}=0, \quad D_{\phi}v^{(p)} - d_{A}^{H}v_{(p-1)}=0 \label{zero-modes}
\ee

It is
straightforward to see that if $D_{\phi}$ is invertible on
$\Omega_{H}^{p}(M,E)\oplus \Omega_{H}^{p-1}(M,E)$ (so that
$\Omega_{\parallel}^{p}(M,E)\oplus \Omega_{\parallel}^{p-1}(M,E)$ is
trivial) then the de Rham group $\mathrm{H}_{\mathcal{A}}^{p}(M,E)$ is
trivial. Let $u$ be a representative of $[u] \in
\mathrm{H}_{\mathcal{A}}^{p}(M,E)$ then $v = u +
d_{\mathcal{A}}\lambda$ is also a representative of $[u]$. If
$\Omega_{\parallel}^{p-1}(M,E)=0$ then one may choose $\lambda$ so
that $v_{(p-1)}=0$. The condition that $d_{\mathcal{A}}v=0$ (\ref{zero-modes}) now
includes the equation $D_{\phi}v^{(p)}=0$, but by assumption there are
no solutions to this so that $v=0$ and consequently
$\mathrm{H}_{\mathcal{A}}^{p}(M,E)=0$.

More generally, when $D_{\phi}$ has a kernel pick a $v$ so that
$v_{(p-1)}^{\perp}=0$. The last equation in (\ref{zero-modes}) now
breaks into two parts
\be
D_{\phi}v^{(p)}_{\perp}=0, \quad d_{A}^{H}v_{(p-1)}^{\parallel}=0
\ee
of which the first, by definition, implies that
$v^{(p)}_{\perp}=0$. Consequently all the cohomology lives in the
parallel fields so that the zero mode measure is a part of
$Z_{\parallel}[A]$. Indeed in the
gauges chosen there are two types of zero modes, namely
either
\be
  d_{A}^{H}v^{(p)}_{\parallel}=0 \quad, \quad v_{(p-1)}^{\parallel}=0 
\ee
or $v_{(p-1)}^{\parallel}\neq 0$ and
\be 
  d_{A}^{H}v_{(p-1)}^{\parallel}=0 \quad, \quad d_{A}^{H}v^{(p)}_{\parallel}=- d\kappa
    v_{(p-1)}^{\parallel}
    \ee
The second of these says that any harmonic $v_{(p-1)}^{\parallel}$
such that $d\kappa \wedge v_{(p-1)}^{\parallel}$ is equal to a standard
part of the gauge symmetry is a zero mode. In short, all zero modes
are harmonic but the converse need not be true.

Turning back to the discussion around (\ref{cov-gf}) those harmonic
$\Lambda_{(p-2)}^{\parallel} $ for which $d\kappa \wedge
\Lambda_{(p-2)}^{\parallel} $ are of the form $d_{A}^{H}
\sigma^{(p)}_{\parallel}$ are such that one can redefine
$\Lambda^{(p)}_{\parallel}$ in (\ref{parallel-trans}) to eliminate the
term proportional to $d\kappa$ from the transformations. Such harmonic
$\Lambda_{(p-2)}^{\parallel}$ are zero modes of the theory. They do not
appear in the action nor in the transformations. The second type of
harmonic $\Lambda_{(p-2)}^{\parallel} $ are those for which $d\kappa \wedge
\Lambda_{(p-2)}^{\parallel} $ is harmonic, hence cannot be expressed
as $d_{A}^{H}
\sigma^{(p)}_{\parallel}$ and so cannot be eliminated from the gauge
symmetry (\ref{parallel-trans}) transformations. Such
$\Lambda_{(p-2)}^{\parallel} $ are part of the gauge group. They are
harmonic but they are not zero modes. These modes are used
to gauge fix some of the harmonic
$\Lambda^{(p+1)}_{\parallel} $ modes which have not been used to impose the
covariant conditions (\ref{cov-gf}). In Example \ref{example-zm} we
show how this is done in practice. In summary some of the harmonic
modes are part of the gauge symmetry and are to be gauge fixed in the
standard BRST fashion. Those which are not part of the gauge symmetry
are zero modes and need to be dealt with in a different manner.

How to
deal with zero modes in the path integral representation has been
described in detail in
\cite{BT-BF,BKT}. In particular, it has been shown in \cite{BKT} how a choice of
gauge matches the definition of Ray and
Singer \cite{Ray-Singer-Analytic}. We will not repeat that analysis
here, but rather take the attitude that the zero modes have been
dealt with according to one of the above prescriptions. Indeed
we take the stronger attitude that all harmonic modes have been dealt
with and in this way $\mu$ has been determined.

Before leaving this section, however, we pose a question. Could the
(horizontal and parallel) ``torsion'' $\tau(A)$ by itself be an
interesting invariant of the base space $N$? From the viewpoint of
the path integral, varying the metric on $N$ amounts to a BRST exact
term, just as for the standard Ray-Singer torsion, and hence formally
$\tau(A)$ would be invariant under such deformations.

\section{A Formal Expression for
\texorpdfstring{$\boldsymbol{Z_{\perp}[\phi]}$}{Z[phi]}}\label{section5}

Apart from the zero modes, the complete partition function is just
$Z_{\perp}[\phi]$. That is the complete flat connection on $E$ over
$M$ is not needed - rather it is just the component $\phi$ in the fibre
direction that appears. In a sense $\phi$ contains almost all of the gauge
invariant data of the bundle as we will see when we analyse the
flatness equations in more detail.

To perform the path integral in the algebraic gauges, we can rely on
the discussion in Appendix \ref{appendix2} which shows that the ghost
triangle for the perpendicular fields can be disentangled from the
parallel fields (discussed in the previous section). Furthermore
we only need the right hand edge of the BV-triangle thanks to the
algebraic conditions.

The total action for the path integral of interest  (including the
ghost terms and taking the algebraic conditions into account) is then
\be
S_{\perp} = i\int_{M} \left(\langle B^{(p)}_{\perp}, \, \kappa D_{\phi}
        C^{(q)}_{\perp} \rangle + \kappa \sum_{r=0}^{p-1}
\langle \overline{\omega}^{(r)}_{\perp} , \, *_{H} D_{\phi}
\omega^{(r)}_{\perp} \rangle  + \kappa \sum_{s=0}^{q-1}
\langle \overline{\lambda}^{(s)}_{\perp} , \, *_{H} D_{\phi}
\lambda^{(s)}_{\perp} \rangle  \right) \label{SIAction}
\ee
where the $(\omega_{\perp} , \overline{\omega}_{\perp} )$ are the ghosts for $B_{\perp} $ and
$(\lambda_{\perp} , \overline{\lambda }_{\perp} )$ are the ghosts for $C_{\perp} $ all in the
appropriate representations. By making use
of the Hodge star operator (\ref{hodgeH}) we have
\begin{align}
\int_{M}\kappa \wedge \langle C^{(q)}_{\perp}  , \, D_{\phi}
  B^{(p)}_{\perp}  \rangle  &=  \int_{M}\kappa \wedge \langle
                              \overline{B}^{(p)}_{\perp} , \,  *_{H}  D_{\phi}
    B^{(p)}_{\perp}  \rangle
\end{align}
where $\overline{B}^{(p)}_{\perp} = *_{H}C^{(q)}_{\perp} $ and $\kappa \wedge
*_{H}C^{(q)}_{\perp}  = * C^{(q)}_{\perp} $. By virtue of (\ref{L-star}) $D_{\phi}$
commutes with $*_{H}$ in the integrals. Now all the terms in (\ref{SIAction}) have
the same form.

At least formally, we then
find that the partition function 
takes the form of an infinite-dimensional determinant, 
\be
Z_{\perp}[\phi] = \Det{D_{\phi}}|_{\Omega_{H}^{\perp}(M,E)}
  \ee
where
$\Omega_{H}^{\perp}(M,E)$ is the direct (signed) sum of infinite-dimensional vector spaces 
\begin{align}
\Omega_{H}^{\perp}(M,E) & =  \sum_{i=0}^{p}
(-1)^{p-i+1}\Omega_{\perp}^{i}(M, E) +
                          \sum_{j=0}^{q-1}(-1)^{q-j+1}\Omega_{\perp}^{j}(M,
                          E) \nonumber \\
  & =  \sum_{i=0}^{p}
(-1)^{p-i+1}\Omega_{\perp}^{i}(M, E) +
                          \sum_{j=0}^{n-p-1}(-1)^{n-p-j+1}\Omega_{\perp}^{n-j}(M,
    E) \nonumber\\
  & =  (-1)^{p+1} \sum_{i=0}^{n}
(-1)^{i}\Omega_{\perp}^{i}(M, E) 
\end{align}
(here $(-1)^{2r} $ means $\oplus$ while $(-1)^{2r+1}$ stands for $\ominus$).
Note that the second equality holds by Hodge duality $*_{H}$ so that $\Omega_{\perp}^{j}(M,
    E) \simeq\Omega_{\perp}^{n-j}(M,
    E)$ and that $q=n-p$ while the third equality comes on
    setting $j-n=i$. Of course, in order to define the theory we still need to 
    specify the regularisation that will be used.  All in all the last equality
    then suggests that the
    sum over infinite dimensional vector spaces may reduce to a sum over some finite
    dimensional cohomology spaces. We cannot say more at this level of
    generality. However, once we have specified the flat bundles
    and the flat  connections on them that we are interested in, 
    the formulae we find are indeed cohomological in nature.

\section{Flat Connections on a Pull Back Bundle 
\texorpdfstring{$\boldsymbol{E}=\boldsymbol{\pi}^{*}\boldsymbol{V}$}{E=pi*V}}\label{section6}

In previous sections we have developed the path integral for the
partition function as far as possible without specifying either the vector bundle $E$ or the
allowed flat connection $\mathcal{A}$ on it. In this section we do not
aim for complete generality but, rather, choose a natural class of bundles and
connections.

Recall that the flatness equations $F_{\mathcal{A}}=0$ by (\ref{flat-1}) are
\be
\begin{aligned}
 & F_{A}^{H} = - d\kappa \, \phi \\
 & \mathcal{L}_{\xi} A = d_{A}^{H} \phi \label{flat}
\end{aligned}
\ee
As a first example consider a solution with $\phi=0$. In this case the
solutions satisfy $\mathcal{L}_{\xi} A =0$ and the connection $A$ can
be thought of as a connection on a bundle on $N$. The first equation
of (\ref{flat}) then tells us that the connection $A$ is a flat
connection. So the upshot is that in this case the bundle $E$ is the
pullback of a flat vector bundle on $N$ with flat connection on $E$
being the pullback flat connection $\mathcal{A}= \pi^{*} A$. Of
course in the case at hand, from the results of the previous sections,
the partition function is given completely by $Z_{||}[A]$ and only contains
information about the zero modes, the flat connection $A$ playing no other
role.

A more general and more interesting situation, and the one that we concentrate on here, is
where we still have a pullback bundle but with $\phi \neq 0$. Let
$V\rightarrow N$ be a $G$ vector bundle over $N$. $V$ is taken to 
be such that it admits a connection $A$ which satisfies
\be
\begin{aligned}
 & F_{A}= - d\kappa \, \phi \\
 & d_{A}\phi =0\label{flat-N}
\end{aligned}
\ee
where $\phi \in \Omega^{0}(N, \ad P_{V})$ and
$P_{V}$ is the associated principle bundle for $\phi \neq 0$. By
construction one admits a $\phi$ such that $\mathcal{L}_{\xi}\phi=0$
from the point of view of $M$.

Now fix the bundle $E= \pi^{*}V$ and equip it with the connection
$\mathcal{A}= A + \kappa \phi$. This is not the pullback connection
$\pi^{*}A \simeq A$ which is not flat. However, as one can add to any
connection a Lie algebra valued 1-form and still have a connection,
we have added $\kappa \phi$ to $A$ and,  by construction, we now have
$\mathcal{A}=A + \kappa \phi$. This connection is a flat connection on the pullback
bundle as (\ref{flat}) is equivalent to (\ref{flat-N}) for these fields.

We fix on $G=SU(r)$. Solutions to the equation $d_{A}v =0$ with $v \in
\Omega^{0}(N, \, \ad P_{V})$
span the Lie
algebra of the stabiliser of the connection $A$ and a connection with
non-trivial stabiliser is reducible. The
first part of the flatness equation tells us that the curvature of $A$
is non-zero only in the Abelian direction generated by $\phi$. In this
way we see that the connections of interest reduce to the form
$K\times U(1)$ and are flat in the $K$ direction. The connection can be expressed as $A^{SU(r)} =
A^{K} \oplus A^{U(1)} $ with curvature 2-form that satisfies
\be
F_{A^{K}}=0, \quad F_{A^{U(1)}} = - d\kappa \, \phi \label{flat-F}
\ee
With such a split, one has $[A, \phi]=0$  so that the second equation of
(\ref{flat-N}) together with $\mathcal{L}_{\xi}\phi=0$ implies that
$d\phi=0$ on $M$, i.e.\ that $\phi$ is constant. Note that not only
the connection $A$ is reducible, but also the flat connection $\mathcal{A}=
A + \kappa \phi$ is, since 
\be
d_A\phi = 0 \quad\Rightarrow\quad d_{\mathcal{A}}\phi=0\;\;.
\ee
Note that, with $\phi$ constant, the curvature formula 
\be
F_A = -\phi d\kappa = d (-\phi\kappa) 
\ee
shows that all of the Chern-Weil representatives of the characteristic classes
of the pullback bundle $\pi^{*}V$ on $M$ are trivial. However, the bundle $V$ 
on $N$ may well be non-trivial. After all, on $N$ one has $c_1(\mathcal{L}_M) =
[d\kappa]$. This will be seen in the examples below. 

In the following, in order to illustrate the evaluation of the path
integral, we will not consider the most general reducible connection
but restrict our attention to reducible connections with $A^{K}=0$
(though the more general case can also be dealt with), and we write
$A^{U(1)}=A$. Moreover,
since $\phi$ is constant we may as well take it to be diagonal. From
here on we do not carry around the $U(1)$ superscript as it is
clear  which component of $A$ is non-zero.

Let $d\kappa$ correspond to $p_{M}$ times an integral class on $N$ (for details
see Appendix \ref{appendix1}). Then (\ref{flat-F}) makes sense for
$\phi$ having any entries (as a matrix) of the form
$2\pi i b_{j}/p_{M}$ with the $b_{j}$ integral and $\sum_{j=1}^{r}b_{j}=0$. The non-trivial
equations for $A$ in (\ref{flat-F}) with $G$
simply connected and $I(G)$ the integral lattice of $G$ then imply that,
\be
\phi \in I(G)/p_{M} \label{phi-pm}
\ee

For example, with $G=SU(2)$, we have
\be
\phi = 2\pi i n/p_{M}
\begin{pmatrix}
         1 & 0 \\
         0 &-1
\end{pmatrix}
\quad n \in \mathbb{Z}_{\neq 0} \label{phi-n}
\ee
By $F_A = -d\kappa\phi$, 
a $\phi$ as in (\ref{phi-n}) implies that one has a non-trivial
complex line bundle $\mathcal{L}^{n}$ on
$N$ with $U(1)$ connection $A$ on $\mathcal{L}$ having integral real first Chern class
(cf.\ \eqref{achern1})  and \eqref{aomegam}) 
\be
c_{1}[\mathcal{L}]= [iF_{A}/2\pi] = 1/p_{M}\, [\omega_M] \equiv [\omega]
\ee
Note that $\mathcal{L}_{M}= \mathcal{L}^{p_{M}}$ where
$\mathcal{L}_{M}$ is the line bundle associated to the circle bundle
which is $M$. The $SU(2)$
vector bundle $V\rightarrow N$ defined by (\ref{phi-n}) is split according to
\be
V= \mathcal{L}^{n} \oplus \mathcal{L}^{- n}
\ee
with total Chern class
\be
c(V) = 1 - n^{2} c_{1}(\mathcal{L})\wedge c_{1}(\mathcal{L})
\ee
Likewise for the complexified adjoint bundle
\be
\ad P_{\mathbb{C}} = \mathcal{L}^{0} \oplus
\left(\mathcal{L}^{2n}\oplus \mathcal{L}^{-2n} \right), \quad c(\ad 
P_{\mathbb{C}}) = 1 -4n^{2} c_{1}(\mathcal{L})\wedge c_{1}(\mathcal{L}) \label{su2-line}
\ee
Consequently both $V$ and $\ad P_{\mathbb{C}}$ may well be non-trivial. For example on $N= \mathbb{P}^{1}$ these are obviously
trivial while on $N= \mathbb{P}^{m}$ for $m\geq 2$ they are not.

For $G=SU(r)$, given our choices for the flat connections, the vector
bundle $V$ for a general representation is taken to split as
\be
V = \bigoplus_{i=1}^{k}\; \mathcal{L}^{q_{i}} , \label{Vlines}
\ee
and fields are direct sums of sections of $\Omega^{p}(N,
\mathcal{L}^{q_{i}} \otimes \mathcal{L}^{r} _{M})$ for $r \in
\mathbf{Z}$ and $i=1, \dots, k$ (see Fact \ref{fact4} in Appendix \ref{appendix1}).  Note that for $\omega_{r,q}^{(p)} \in \Omega^{p}(N,
\mathcal{L}^{q_{i}} \otimes \mathcal{L}^{r} _{M})$
\be
D_{\phi}\, \omega_{r,q}^{(p)}=(2\pi i r +  2\pi i q/p_{M})\, \omega_{r,q}^{(p)}
\ee
and we see that zero modes occur precisely when $\mathcal{L}^{q}=
\mathcal{L}_{M}^{-r}$, and the eigenvalues are independent of the form degree of
$\omega_{r,q}^{(p)} $. 

The determinant of the $D_{\phi}$ on $\Omega^{p}_{H}(M,
\pi^{*}\mathcal{L}^{q} )$ is
\be
\Det{D_{\phi}}|_{\Omega^{p}_{H}(M,
  \pi^{*}\mathcal{L}^{q} )} \simeq \prod_{r \in \mathbf{Z}}(2\pi i r +
2\pi i q/p_{M})|_{\Omega^{p}(N,
\mathcal{L}^{q} \otimes \mathcal{L}^{r} _{M})}\label{DetD}
\ee
Therefore, if $q$ is an integer multiple of $p_M$, $q=0 \bmod p_M$, 
there is a zero mode eigenvalue. In particular, if $p_{M}=\pm 1$ then such
zero modes are unavoidable in any representation. By the same logic
the determinant (\ref{DetD}) does not depend on $q$ but rather on $ q\,
\mathrm{mod}\, p_{M}$. The zero mode in (\ref{DetD}) belongs in $Z_{\parallel}$ and is
to be deleted in the calculation for $Z_{\perp}$. By shifting $r$ if
required, the contribution to $Z_{\perp}$ of the determinant 
(\ref{DetD}) for $q = 0 \bmod p_M$ is
\be
q = 0 \bmod p_M \quad\Ra\quad 
\left(\Det{D_{\phi}}|_{\Omega^{p}_{H}(M,
  \pi^{*}\mathcal{L}^{q} )}\right)_{\perp} \simeq 
\prod_{r \in \mathbf{Z}_{\neq 0}}(2\pi i r )|_{\Omega^{p}(N, \mathcal{L}^{r} _{M})} \label{norms}
\ee
Though strictly a contribution to $Z_{\perp}$ we will move such normalisation
terms (\ref{norms}) to $Z_{\parallel}$; their value can be determined from the
calculation of the following section.

With the above understood, we now decompose the vector bundle $V$ given
in (\ref{Vlines}) as
\be
\begin{aligned}
&  V  = V_{\perp}\bigoplus V_{\parallel} & \\
& V_{\perp}  = \bigoplus_{i=1}^{k_{1}}\,
                  \mathcal{L}^{q_{i} } \quad &\text{for}\;
              q_{i} \neq 0 \bmod\, p_{M} \\
&   V_{\parallel}  = \bigoplus_{i=k_{1}+1} ^{k}\,
                  \mathcal{L}^{q_{i}}  \quad &\text{for}\; 
                   q_{i} = 0 \bmod p_{M}\label{Vdecomp}
\end{aligned}
\ee
and obtain the bundles 
\be
E_{\parallel} = \pi^{*}V_{\parallel}, \quad E_{\perp} = \pi^{*}
V_{\perp} \label{Edecomp}
\ee
on $M$, with $E_\perp$ the bundle that will contribute to
$Z_{\perp}[\phi]$. Since the eigenvalues of $D_{\phi}$
are independent of the form degree, the decomposition
(\ref{par-perp}) is thus more explicitly given by
\be
\Omega_{\parallel}^{p}(M, E) = \Omega_{H}^{p}(M, E_{\parallel}), \quad
\Omega_{\perp}^{p}(M, E) = \Omega_{H}^{p}(M, E_{\perp})
\ee

Before turning to the actual calculation (section \ref{section7}), 
we present some examples of the bundles under consideration.

\begin{example}{\rm{Let us consider the adjoint representation. The
      complexified Lie algebra splits as 
\[
\lg_{\mathbb{C}}= \lt_{\mathbb{C}} \oplus_{\alpha \in
    \Delta_{+}} \left(V_{\alpha} 
    \oplus V_{-\alpha}\right)
  \]
where the root spaces $V_{\alpha}$ are copies of
$\mathbb{C}$ and the space of positive roots is denoted by
$\Delta_{+}$.  Just as for the $SU(2)$ example (\ref{su2-line}), for a
semi-simple group $G$ the $\ad P_{\mathbb{C}}$ bundle over $N$ is
split into copies of line bundles corresponding to the positive and negative roots spaces 
\[
\ad P_{\mathbb{C}} = \left(\oplus_{\dim{\lt}}\,\mathcal{L}^{0}\right)
\oplus \oplus_{\alpha \in \Delta_{+}} \left(\mathcal{L}_{M}^{-\alpha(i\phi)}
  \oplus \mathcal{L}_{M}^{\alpha(i\phi)}\right) 
\]
Here we are taking a liberty in the notation as $\alpha(i\phi)$ may be
fractional. One should understand $\mathcal{L}_{M}^{\pm
  \alpha(i\phi)} \equiv \mathcal{L}^{\pm
  p_{M}\alpha(i\phi)}$.
The
covariant derivative (\ref{cov-time-der}) acting on sections is
\[
D_{\phi}\, \omega^{(p)}_{r\, \alpha} = (2\pi i r + \alpha(\phi))
\omega^{(p)}_{r\, \alpha} , \quad \omega^{(p)}_{r\, \alpha} \in
\Omega^{p}(N, \, \mathcal{L}_{M}^{-\alpha(i\phi)}  
\otimes \mathcal{L}_{M}^{r}) 
\]
    }}
  \end{example}

\begin{example}{\rm{Let $M$ be the Lens $L_{2r+1}(p)$ that is the $S^{1}$ bundle
      $S^{1}\rightarrow L_{2r+1}(p) \rightarrow \mathbb{CP}^{r}$ with
      $d\kappa=p\, \pi^{*}\omega$ where $\omega$ is the K\"{a}hler 
      2-form dual to the integral homology 2-cycle of
      $\mathbb{CP}^{r}$. The only flat connection on
      $\mathbb{CP}^{r}$, up to gauge equivalence,
      is the trivial connection  (as is true on all simply connected
      manifolds). This means that one may
      concentrate on the non-trivial $U(1)$ connection. The Abelian 
      Ray-Singer Torsion has been computed for these spaces by Ray
      \cite{Ray} in order to corroborate the conjecture of Ray and Singer
      that their torsion coincides with Reidemeister Torsion. We
      re-derive this result in Example \ref{ex:lens} below. 
    }}
\end{example}

\begin{example}{\rm{Let $(N, \omega)$ be a K\"{a}hler manifold  with dimension
      $n=2r$ and K\"{a}hler 
      2-form $\omega$ where $d\kappa =\pi^{*}\omega$ and take $V$ to
      be a vector bundle on $N$. Solutions to the flatness equations on $M$
      (\ref{flat}), which are constant along the flow
      $\mathcal{L}_{\xi} A=0$, are special solutions of the
      Yang-Mills equations on $N$ (here $d$ is the exterior derivative
      on the base)
      \[
      d_{A}* F_{A} =0 
      \]
      This follows as on a K\"{a}hler manifold $*\omega =
      \omega^{r-1} /(r-1)!$ and the evolution equation together with the gauge condition on
      $\phi$ lead to $d_{A}\phi =0$. 
      This situation is familiar from the study of flat
connections in Chern-Simons theory on contact 3-manifolds by Beasley
and Witten \cite{BW} .}}
\end{example}

\section{Evaluation of the Path Integral Representation of 
\texorpdfstring{$\boldsymbol{Z_{\perp}[\phi]}$}{Z[phi]}}\label{section7}

Summarising the previous sections we note that (\ref{flat-F}) tells us
which line bundles one is considering and, as 
$\kappa \phi$ is the vertical component of the connection, it also determines
the holonomy along the fibre of the $S^{1}$ bundle. What is missing
is the gauge invariant data in the possible holonomies of $A$ but it
seems that the Ray-Singer torsion does not see this data for these
backgrounds. In particular the information about $A$ is contained in
$Z_{\parallel}$ and so enters through the zero mode measure $\mu$ in
(\ref{zero-mode-mu}). Apart from the zero mode sector the Ray-Singer
torsion is completely determined by $Z_{\perp}[\phi]$ whose evaluation
we now turn to.

The perpendicular fields that enter into the action  (\ref{SIAction}) take values in
$E_{\perp}$, the pullback of $V_{\perp}$ as defined in
(\ref{Vdecomp}, \ref{Edecomp}). The partition function of the action (\ref{SIAction}) is given by
products of path integrals over forms in $\Omega^{p}_{H}(M,
\pi^{*}\mathcal{L}^{q})$ and its dual for $q \neq 0 \, \mathrm{mod}\, p_{M}$ of the form
\be
\int D \Phi^{(p)}_{q}D \overline{\Phi}^{(p)}_{-q} \,
  \exp{\left(i\int_{M}\kappa \wedge \langle
      \overline{\Phi}^{(p)}_{-q}, \, *_{H} D_{\phi}
   \Phi^{(p)}_{q}\rangle \right)}
 = (\Det{D_{\phi}})^{\pm1}|_{\Omega^{p}_{H}(M, \pi^{*}\mathcal{L}^{q}
   ) } 
\ee
which, using \eqref{DetD}, can be written more explicitly as
\be
 (\Det{D_{\phi}})^{\pm1}|_{\Omega^{p}_{H}(M, \pi^{*}\mathcal{L}^{q}
   ) } 
 = \prod_{r \in\mathbb{Z}} (2\pi i r + 2\pi i q/p_{M}
   )^{\pm 1}|_{\Omega^{p}(N, \mathcal{L}^{q} \otimes
   \mathcal{L}_{M}^{r}) } \label{PI-det}
\ee
The sign in the exponents depends on
the statistics of the fields being integrated over. 
The notation in
(\ref{PI-det}) indicates that one gets the indicated products on the 
infinite-dimensional spaces 
$\Omega^{p}(N, \mathcal{L}^{q} \bigotimes
   \mathcal{L}_{M}^{r})$, so these 
clearly still require some form of
   regularisation.
 We adopt a Heat Kernel regularisation where the
   Laplacian $\Delta_{A}= d_{A}\, d_{A}^{*} + d_{A}^{*}\, d_{A} $ that
   is used is that based on the twisted de Rham 
   operator $d_{A}$ on $N$ and the Hodge operator is $*_{H}$. The
   infinite product over the Fourier modes also 
   requires regularisation. As both types of regularisation have been
   explained in detail in the 3 
   dimensional case \cite{btg/g, BW, btseifert}, and more generally in
   higher dimensions \cite{BKT} and as there are no extra subtleties in
   the present situation, we skip the details here.

   There is one aspect of the path integral evaluation that we do need
   to comment on, however. The integrals are all of delta function
   type here
   \be
   \int \frac{dx}{\sqrt{2\pi}}\frac{dy}{\sqrt{2\pi}}\, \exp{\left(ix.D.y\right)} = | \Det{D}|^{-1}
   \ee
  (with a similar conclusion for the Grassman odd integrals) so that
  there is no phase. In Section \ref{section8} we give
  some examples where, just as in
  Chern-Simons theory \cite{WCS}, there is a phase and it needs to
  be determined. We do not show the absolute value symbols below,
  however, they are to be understood up to and including (\ref{RSTperp}).

Up to metric-dependent factors (which also appear in $Z_{\parallel}$), which drop out of
the complete path integral, the partition function is essentially
\begin{multline}
Z_{\perp}[\phi] \simeq \\
\prod_{i=1}^{k_{1}}\prod_{r \in\mathbb{Z}} \left(\prod_{s=0}^{p}
\left(\Det{D_{\phi}}|_{\Omega^{s}(N, \,  \mathcal{L}^{
                         (q_{i}-rp_{M})})} \right)^{(-1)^{s+p+1}}
  \prod_{t=0}^{q-1}
\left( \Det{D_{\phi}}|_{\Omega^{t}(N, \, \mathcal{L}^{ (q_{i}-rp_{M})})}\right)
^{{(-1)^{t+q+1}}}\right) \label{prodZ} 
\end{multline}
By making use of the fact that for a vector bundle $V$ on $N$ that
$\Omega^{t}(N,V )\simeq\Omega^{n-t}(N, V^{*})$ for fixed $r$ and $q_{i}$ 
as well as that $n-q+1 = p+1$
(and on recalling that it is the absolute value of the determinant that
appears in the second term in
(\ref{prodZ})) one has
\be
Z_{\perp}[\phi]\simeq   \prod_{i=1}^{k_{1}}\prod_{r \in\mathbb{Z}}
\prod_{s=0}^{n} \left( (\Det{D_{\phi}})|_{\Omega^{s}(N,
    \mathcal{L}^{ (q_{i}-rp_{M})} )}\right)^{(-1)^{s+p+1}} \label{simplified}
\ee
Suitably regularised, the alternating sum of the infinite dimensional vector spaces
of differential forms 
reduces to the alternating sum of finite dimensional
spaces of harmonic forms. This comes about as follows: while the  eigenvalues of the
$D_{\phi}$ operator do not depend on the form degree their multiplicity
does, so by fixing on each $r$ and $q_{i}$ one is calculating
\be
\ln{Z_{\perp}[\phi]} \simeq  (-1)^{p+1}\sum_{i=1}^{k_{1}}\sum_{r \in\mathbb{Z}}\left[
 \ln{(2\pi i r + 2\pi q_{i}/p_{M})}\Tr{\left((-1)^{F}
         \exp{(- \epsilon \Delta_{A})}\right)}\right]
   \ee
   Here the trace is over the space of forms $\bigoplus_{s=0}^{n}\Omega^{s}(N, \mathcal{L}^{
  (q_{i}-rp_{M})}) $, $(-1)^{F}$ is positive on forms of even
degree and negative on forms of odd degree, and we have pulled out
$\ln{D_{\phi}}$ from the trace as its eigenvalues are independent of
the form degree $s$. One recognises the
trace as the Witten index of the twisted de Rham operator $d_{A}$ and
that trace reduces to one of $(-1)^{F}$ over the Hodge groups $\bigoplus_{s=0}^{n}\mathrm{H}^{s}(N, 
\mathcal{L}^{(q_{i}-rp_{M})} ) $.  The index of the twisted de Rham
operator (Theorem 3.4.3 in \cite{Gilkey}) is just the Euler
characteristic $\chi(N)$ of the underlying manifold 
and does not depend on the line bundles $\mathcal{L}^{ (q_{i}-rp_{M})}$.

Reassembling all the pieces of the path integral the partition
function is
\be
Z_{\perp}[\phi]\simeq   \prod_{i=1}^{k_{1}}\prod_{r \in\mathbb{Z}}\left(
  2\pi i r + 2\pi i q_{i}/p_{M}\right)^{(-1)^{p+1} \chi(N)} \label{RSTperp}
\ee
Note that the infinite product is an expansion of the hyperbolic
$\sinh$ function
\be
\prod_{r \in\mathbb{Z}}\left(
  2\pi i r + X\right) = 2 \sinh{(X/2)} \label{sinh}
\ee
The Ray-Singer Torsion for a line
bundle $\mathcal{L}$  over the circle is
(see \cite{Ray})
\be
\tau_{S^{1}}(\phi, \, \mathcal{L}) =  2 |\sinh{( \phi/2)}|
\ee
with $\phi$ the anti-Hermitian connection on $\mathcal{L}$. Ultimately
then the path integral evaluates to 
\be
Z_{\perp}[\phi] = \left(\tau_{S^{1}}(\phi, E_{\perp}|_{S^{1}})^{\chi(N)}\right)^{(-1)^{p+1}}
\ee
where $E_{\perp}|_{S^{1}}$ is the restriction to the fibre of the
bundle $E_{\perp}$ over $M$ (recall that $\phi$ does not depend on the
base point of the fibre on $N$).

Taking into account the contribution of $Z_{\parallel}[A]$ we find that
(up to a suitable normalisation) the complete partition function takes 
the form
\be
Z[\mathcal{A},E] = \mu\,  . \, \left(\tau_{S^{1}}(\phi, E_{\perp}|_{S^{1}})^{\chi(N)}\right)^{(-1)^{p+1}}
\ee
Given the relationship between the partition function and the
Ray-Singer torsion (\ref{part-tor}) we re-write this as
\be
\tau_{M}(\mathcal{A}, E) = (\mu)^{(-1)^{p+1}}\, .\, \tau_{S^{1}}(\phi,
E_{\perp}|_{S^{1}})^{\chi(N)} \label{RST-Complete}
\ee
where $\mu^{-1}$ is understood to be the appropriate section of the
inverse determinant line. The formula (\ref{RST-Complete}) is our main
result giving a complete evaluation of the path integral
representation of Schwarz for the Ray-Singer Torsion on non-trivial
$S^{1}$ bundles.

In the special case that $E=E_{\perp}$ one has
\be
\tau_{M}(\mathcal{A}, E )= \tau_{S^{1}}(\phi,
E|_{S^{1}})^{\chi(N)} \label{RST-Fried-F}
\ee
The result (\ref{RST-Fried-F}) is a special case of Fried's theorem for the Reidemeister
 torsion (result (V) on page 26 of \cite{Fried}) where the fibre of the bundle that he considers is taken to
 be $S^{1}$ while (\ref{RST-Complete}) is a generalisation of Fried's
 result to non-acyclic representations.

 \begin{example}\label{ex:lens}{\rm{The Lens space obtained by the identification
       $(z_{1}, \dots , z_{r+1}) \sim  \zeta (z_{1}, \dots , z_{r+1})$
       where $\zeta$ is a $p$-th root of unity and $\sum_{j=1}^{r+1}
       |z_{j}|^{2}=1$ is an $S^{1}$ bundle over $\mathbb{CP}^{r}$. Ray
       finds that for these Lens spaces the Abelian
       Ray-Singer Torsion has the explicit form (this follows from
       equation (1) in \cite{Ray})
       \[
       \tau_{L_{2r+1}}(\zeta) = |1- \zeta|^{r+1}=
       \tau_{S^{1}}(\zeta)^{\chi(\mathbb{CP}^{r})} 
       \]
       where $\zeta$ is to be interpreted as the holonomy of the Abelian
       connection on a line bundle $\mathcal{L}$ over
       $S^{1}$. His result follows from (\ref{RSTperp}) on taking $N=
       \mathbb{CP}^{r}$ and $k_{1}=1$.
       
     }}
 \end{example}

\section{Some Generalisations and Observations}\label{section8}

There are various generalisations that may be considered where the
techniques of this paper can apply. We list three of them.
\begin{itemize}
  \item
We looked at the case that $M$ is a smooth $S^1$-bundle over
a smooth manifold $N$. In the 3-dimensional case, the calculation
can readily be extended to Seifert fibred 3-manifold with
base an orbifold of genus $g$. The $i$'th orbifold point has order
$a_{i}$ for $i=1, \dots , 
N$. In this case $Z_{\perp}$ with action $S _{\perp} $
gives a well known result, namely \cite{btseifert}
\be
Z_{\perp}[\phi] = \tau_{S^{1}}(\phi)^{2-N-g} \prod_{i=1}^{N} \tau_{S^{1}}(\phi/a_{i})
\ee
It ought to be possible to perform similar calculations on smooth
manifolds $M$ which are $S^{1}$ bundles over orbifolds in higher
dimensions as well. One would need, as in the 3-dimensional case, to
make use of the Kawasaki index theorem \cite{Kawasaki} in order to deal with the
orbifold points and to have an orbifold version of the Euler
characteristic. A possible application would be to the general Lens
spaces $L_{2r+1}(\nu_{1}, \dots , \nu_{r+1})$ considered by Ray
\cite{Ray} where
$(z_{1}, \dots , z_{r+1}) \sim   (\zeta^{\nu_{1}}z_{1}, \dots ,
\zeta^{\nu_{r+1}}z_{r+1})$ for $\zeta $ a $p$-th root of unity and
$\sum_{i=1}^{r+1} |z_{i}|^{2}=1$ which are $S^{1}$ bundles over weighted
projective spaces.

\item One may also consider the case of non-simply connected groups such as
$PSU(n)\simeq SU(n)/\mathbb{Z}_{n}$. Some of the flat $PSU(n)$ bundles
will be flat $SU(n)$ bundles so that the results carry over. However,
not all $PSU(n)$ bundles arise from $SU(n)$ bundles as can be seen for
the $SU(2)$ case in (\ref{su2-line}) 
where the adjoint bundle only allows even powers of line bundles while
$SO(3)$ bundles would include odd powers (which would, in turn, correspond to
having non-trivial Stiefel-Whitney classes). It is possible to envisage being
able to use the results of the body of this paper by taking $n$ in (\ref{su2-line}) to be half
integral, though such an approach requires further investigation.

\item As explained in \cite{BT-BF}  (pages 152-153) if $M$ has real dimension $m=4k-1$
  then one may refine the theory and  consider an action of the form
\be
S = \int \langle B , \,  d_{\mathcal{A}}B\rangle
\ee
where $B \in \Omega^{2k-1}(M, \lg)$. This path integral leads to the
square root of the Ray-Singer Torsion up to a phase. The phase has
the same source in the path integral as the familiar framing anomaly in
Chern-Simons theory \cite{WCS} and it ought to be possible to get
explicit formulae for the phase for higher dimensional $S^{1}$
bundles. In case $M$ has dimension $m=4k + 1$ 
an appropriate action is 
\be
S = \int \langle \Psi, \,  d_{\mathcal{A}}\Psi\rangle
\ee
where $\Psi\in \Omega^{2k}(M, \lg)$ and Grassman odd.
\end{itemize}

\subsection*{Acknowledgements}

M.\ Kakona thanks the External Activities Unit of the ICTP for a
Ph.D. grant at EAIFR in the University of Rwanda and the HECAP group 
at the   ICTP for supporting this work. 
The work of M.\ Blau is supported by the NCCR SwissMAP (The Mathematics
of Physics) of the Swiss Science Foundation.

\appendix

\section{Some Background on the Geometry of \texorpdfstring{$S^{1}$}{S1} Bundles}\label{appendix1}

Let $\pi: M \to N$ be a smooth $S^1$-bundle over the smooth manifold $M$,
and denote by $\xi$ the fundamental nowhere vanishing vector field generating the
$S^1$-action on $M$. Writing a connection on this bundle as $(-2\pi i)\kappa$ 
for $\kappa$ a globally-defined 1-form on $M$,  the condition that this is 
a connection is equivalent to requiring that $\kappa$ satisfies
\be
\iota_\xi \kappa = 1 \quad,\quad \mathcal{L}_\xi \kappa = 0 \quad\Ra\quad \iota_\xi d\kappa = 0
\label{akappa1}
\ee
where
\be
\mathcal{L}_{\xi} = \iota_{\xi} d + d \iota_{\xi}
\ee
is the Lie derivative. We choose one such $\kappa$ in the following.

As always, the extra structure provided by a choice of connection
provides one with a notion of horizontality and, concretely, in the 
case at hand allows one to create corresponding projection operators
\be
P_{H}= \iota_{\xi} \,  \kappa , \quad P_{V}= \kappa \, \iota_{\xi}
\ee
on differential forms. One easily checks that indeed
\be
P_{H}+ P_{V} = 1, \quad P_{H}^{2}=P_{H}, \quad P_{V}^{2}=P_{V}, \quad
P_{H}P_{V}=P_{V}P_{H} =0
\ee
as required for projectors. 
Consequently, one can decompose differential forms $\alpha \in \Omega^{p}(M,
\bullet)$ as
\be
\alpha= \alpha^{(p)}+ \kappa  \, \alpha_{(p-1)} = P_{H}.\alpha +
P_{V}.\alpha \label{form-split}
\ee
where both ``components'' are horizontal, i.e.\ 
\be
\alpha^{(p)} = \iota_\xi (\kappa \wedge \alpha) \;\Ra\;\iota_{\xi} \alpha^{(p)}=0\quad, \quad 
\alpha_{(p-1)} = \iota_{\xi}\alpha \;\Ra\; \iota_{\xi}\alpha_{(p-1)} = 0 
\ee
One denotes the horizontal spaces of forms by $\Omega_{H}^{p}(M)$ and,
by the decomposition (\ref{form-split}), one has naturally
\be
\Omega^{p}(M) = P_{H}\Omega^{p}(M) \oplus P_{V}\Omega^{p}(M) \simeq
\Omega_{H}^{p}(M) \oplus \Omega_{H}^{p-1}(M) 
\ee
Likewise, one may project the action of the de Rham operator $d:
\Omega^{p}(M)  \rightarrow \Omega^{p+1}(M) $ as follows
\be
d= d_{H}+ \kappa\, D, \quad d_{H} = P_{H}\, d=
\iota_{\xi}\kappa d, \quad \kappa D = P_{V}\, d, \quad 
D= \iota_{\xi} d \label{split-deriv}
\ee
where
\be
d_{H}: \Omega_{H}^{p}(M) \rightarrow \Omega_{H}^{p+1}(M), \quad D:
\Omega_{H}^{p}(M) \rightarrow \Omega_{H}^{p}(M)
\ee
The horizontal de Rham operator on horizontal forms may be written as
\be
d_{H}= d - \kappa \, \mathcal{L}_{\xi} \label{dHd}
\ee
and on horizontal forms the vertical part of the exterior derivative
is the Lie derivative along the $S^{1}$ vector field $\xi$
\be
D = \mathcal{L}_{\xi}, \quad \mathcal{L}_{\xi}
\equiv \iota_{\xi}d + d\iota_{\xi} 
\ee
Unlike the de Rham operator the horizontal de Rham operator
does not square to zero on horizontal forms but rather is proportional
to the Lie derivative with respect to the $S^{1}$ vector field
\be
d_{H}^{2} = -d\kappa\, \mathcal{L}_{\xi}
\ee
Therefore, if one further restricts to forms which are not only horizontal but
for which the Lie derivative vanishes then one has a standard
cohomology called basic cohomology where the horizontal exterior
derivative coincides with the de Rham operator by (\ref{dHd}). Denote the space of differential
forms which are both horizontal and have vanishing Lie derivative by
$\Omega_{\mathrm{Basic}}(M)$. Then one has the following important results: 

\begin{fact}\label{fact1}{\rm{The projection
$\pi:M \rightarrow N$ leads to an isomorphism of de Rham complexes
$\pi^{*}: \Omega(N) \simeq \Omega_{\mathrm{Basic}}(M)$.}}
\end{fact}
This means that basic forms on $M$ are really
pullbacks of forms on the base, so one may view equations involving
them as equations on the base (then pulled back). The isomorphism of complexes
leads to an isomorphism of cohomology groups:
\begin{fact}\label{fact2}{\rm{There is an isomorphism between the
      cohomology groups $H^{q}(N, \, d) \simeq
      H_{\mathrm{Basic}}^{q}(M,  d)$.}}
\end{fact}
In the present context the prime example of a basic closed form 
is provided by $d\kappa$. Indeed, from \eqref{akappa1} one sees
that $\mathcal{L}_\xi \kappa = 0$ (and evidently $d(d\kappa)=0$ 
as well). Therefore, by the above, there exists a closed 2-form 
$\omega_M$ on $N$ such that 
\be
d\kappa = \pi^*\omega_M \quad,\quad d\omega_M = 0\;\;.
\ee
Denoting the associated complex line bundle to the $S^{1}$
bundle by $\mathcal{L}_{M}$, this form $\omega_M$ represents
its 1st Chern class, 
\be
\label{achern1}
c_{1}(\mathcal{L}_{M})= [\omega_M]
\ee
Since this is an integral class, 
there is a largest integer $|p_M|$, $p_M\in\mathbb{Z}$,  such that 
(identifying $H^2(N,\mathbb{Z})$ with its image in $H^2(N,\mathbb{R})$) 
\be
\omega_M =  p_M \omega \quad,\quad [\omega] \in H^2(N,\mathbb{Z}) 
\label{aomegam}
\ee
This integer $p_M$ will enter in the detailed evaluation of the determinants
in the body of the paper. 

Another useful fact is the following:

\begin{fact}\label{fact3}{\rm{Horizontal forms on $M$ have a ``Fourier series'' expansion 
in terms of forms on $N$ as 
\be
\Phi \in \Omega_{H}^{q}(M)\quad\Rightarrow\quad
\Phi = \sum_{m\in \mathbb{Z}} \pi^*\Phi_{m}
\ee
where
\be
\mathcal{L}_{\xi}\Phi_{m} = 2\pi i m\, \Phi_{m}, \quad \Phi_{m} \in
\Omega^{q}(N, \mathcal{L}_{M}^{ m}) 
\ee
}}
\end{fact}

We will also require a metric on $M$ compatible with the $S^{1}$
structure. The metric of choice on $M$ is
$g_{M}$ where
\be
ds^{2}_{M}= \pi^{*}ds^{2}_{N} \oplus \kappa \otimes \kappa \label{gM}
\ee
and $g_{N}$ is a metric on $N$. The associated Hodge
operator commutes with the Lie derivative
\be
\mathcal{L}_{\xi}\circ * = * \circ \mathcal{L}_{\xi} \label{Hodge-Lie}
\ee
as the pullback metric $\pi^{*}ds^{2}_{N}$ is invariant, that is it has $\xi$ as
a Killing vector field, and by
(\ref{akappa1}) the Lie derivative of $\kappa$ vanishes.

Let $\langle \cdot , \cdot \rangle$ be a fixed positive definite fibre
metric on $E$ with which the covariant derivative is compatible.
The Hodge star acts on horizontal forms by
\be
*:\Omega_{H}^{p}(M) = (-1)^{p}\kappa \wedge *_{H}: \Omega_{H}^{p}(M)
\rightarrow \kappa \wedge \Omega_{H}^{n-p}(M) 
\ee
where
\be
*_{H}= \iota_{\xi} \circ (-1)^{p}\, * \label{hodgeH}
\ee
In particular for $\alpha^{(p)}, \, \beta^{(p)} \in \Omega_{H}^{p}(M, E)$
\be
\int_{M} \langle \beta^{(p)} , \, * \alpha^{(p)}\rangle  = \int_{M} \kappa
\wedge \, \langle
\beta^{(p)} , \, *_{H} \alpha^{(p)}\rangle \label{inner-hodge}
\ee
Furthermore,  one has  on horizontal forms that
\be
\mathcal{L}_{\xi} \circ *_{H} = *_{H} \circ \mathcal{L}_{\xi}\label{L-star}
\ee
which one proves as follows
\be
\mathcal{L}_{\xi} *_{H}\, \alpha^{(p)}= (-1)^{p}\mathcal{L}_{\xi} \iota_{\xi}
* \alpha^{(p)}= (-1)^{p}\iota_{\xi} \mathcal{L}_{\xi}  * \alpha^{(p)}=(-1)^{p}
\iota_{\xi} * \mathcal{L}_{\xi}  \alpha^{(p)} = *_{H}\mathcal{L}_{\xi}  \alpha^{(p)} 
\ee
where the first equality follows from the definition (\ref{hodgeH}),
the second as $\mathcal{L} _{\xi}\circ \iota_{\xi} = \iota_{\xi}  \circ
\mathcal{L} _{\xi}$,  the third by (\ref{Hodge-Lie}) and the final
equality once more by the definition (\ref{hodgeH}).

The decompositions of forms (\ref{form-split}) holds for vector bundle
sections over $M$ so that for $E \rightarrow M$
\be
\Omega^{p}(M,E) = P_{H}\Omega^{p}(M,E) \oplus P_{V}\Omega^{p}(M,E) \simeq
\Omega_{H}^{p}(M,E) \oplus \Omega_{H}^{p-1}(M,E) 
\ee

We will be interested in bundles on $M$ which
are pullbacks of bundles over $N$. Recall that the pullback bundle
$\pi^{*} V$ of a bundle $V\rightarrow N$ is such that at each point $y \in
\pi^{-1}(x)$ for $x\in N$ one places a copy of the vector space
$V|_{x}$ so that one has $V|_{x}$ at every point on the fibre above
$x$. Consequently one has the following generalisation of Fact \ref{fact3}.
\begin{fact}\label{fact4}{\rm{Let $V \rightarrow N$ be a vector bundle
      and consider the pullback bundle $\pi^{*}V\rightarrow M$. One
      has the decomposition
      \be
      \Omega^{p}_{H}(M,\pi^{*}V) = \bigoplus_{r \in \mathbb{Z}}
      \Omega^{p}(N,\, V  \otimes \mathcal{L}_{M}^{ r}) 
      \ee
and
$r=0$ corresponds to the basic forms. In case the vector bundle is real one
should consider the decomposition
\be
\Omega^{p}_{H}(M,\pi^{*}V) =  \Omega^{p}(N,\, V ) \bigoplus_{r \in \mathbb{N}_{+}}
      \left( \Omega^{p}(N,\, V  \otimes \mathcal{L}_{M}^{ r}) \oplus
        \Omega^{p}(N,\, V  \otimes \mathcal{L}_{M}^{ -r}) \right)
      \ee
}}
\end{fact}

In particular for $\Phi_{r} \in \Omega^{p}(N, \, V \otimes
\mathcal{L}_{M}^{ r})$
\be
D_{\phi} \Phi_{r} = (2\pi i r + \phi)\Phi_{r} \label{cov-time-der}
\ee
\begin{example}\label{su2-fund}{{\rm Consider the $SU(2)$ case with $V$ the fundamental
      representation bundle $V= M \times \mathbb{C}^{2}$ and $\phi = i \lambda \sigma_{3}$ where
      $\lambda \in \mathbb{R}$. In this case the operator $D_{\phi}$ will
      have zero modes at $\lambda \in 2\pi \mathbb{Z}$, or put another
      way, it will have zero modes when $\exp{(\phi)}$ has eigenvalue
      1. For this bundle, the condition that $\Det(1-
      \exp{(\phi)})\neq 0$ is precisely the acyclicity condition of
      Fried \cite{Fried} mentioned in connection with \eqref{introfried} in the Introduction. 
    }}
\end{example}

\section{Gauge Transformations, Gauge Fixing and BV Triangles}\label{appendix2}

In this appendix we discuss the need for ghosts for ghosts in this theory
and show why one can make do with just the right-hand edge of the
Batalin-Vilkovisky triangle \cite{BV1983} when using algebraic gauges. Furthermore we explain
how the algebraic gauges that we wish to impose are allowed gauge
choices. We do not need to specify the representation of the fields
(the vector bundle $E$).\\

The action (\ref{schwarz}) has the reducible symmetries
\be
 \begin{aligned}
 \delta B  & = d_{\mathcal{A} }\Sigma, \quad \Sigma \in
                       \Omega^{p-1}(M, E^{*})\\
   \delta C  & = d_{\mathcal{A} } \Lambda, \quad \Lambda \in
                         \Omega^{q-1}(M, E) \label{red-symm}
 \end{aligned}
\ee
 These symmetries are reducible since one may vary $\Sigma$ and $\Lambda$
\be
  \begin{aligned}
 \delta \Sigma & = d_{\mathcal{A} }\Psi, \quad \Psi \in
                       \Omega^{p-2}(M, E^{*})\\
   \delta \Lambda  & = d_{\mathcal{A} } \Phi, \quad \Phi \in
                         \Omega^{q-2}(M, E) \label{red-symm-2}
  \end{aligned}
\ee
  as $d_{\mathcal{A} }^{2}=F_\mathcal{A}=0$ without effecting the $B$
  and $C$ transformations (\ref{red-symm}). Likewise one may transform
  both $\Psi$ and $\Phi$ by covariantly exact pieces without changing
  (\ref{red-symm-2}) and so on.

In order to analyse these symmetries it is simplest to deal with
fields that are horizontal. Consequently a 
p-form field $\Phi$ is always decomposed into horizontal fields as
\be
\Phi = \Phi^{(p)} + \kappa \, \Phi_{(p-1)} \label{decomp}
\ee
In accordance with \eqref{split-deriv} and \eqref{dasplit}, we also
decompose the covariant derivative as 
\be
d_{\mathcal{A}} = d_{A} + \kappa \, D_{\phi}, \quad d_{A}^{H} =
\iota_{\xi}\left(\kappa d_{\mathcal{A}}\right) = (1-\kappa
\iota_{\xi})d_{\mathcal{A}}  , \quad D_{\phi} = \iota_{\xi}
d_{\mathcal{A}}
\ee
If the $p$-form $\Phi$ has the usual gauge transformation
\be
\delta \Phi =
d_{\mathcal{A}}\Lambda \label{symm}
\ee
then, with the $(p-1)$-form $\Lambda$ also decomposed  as in
(\ref{decomp}), the gauge symmetry is
\be
\delta \Phi^{(p)} = d_{A}^{H}\Lambda^{(p-1)} + d\kappa \, \Lambda_{(p-2)},
\quad \delta \Phi_{(p-1)} = D_{\phi}\Lambda^{(p-1)} -
d_{A}^{H}\Lambda_{(p-2)} \label{symm2}
\ee
Note the peculiar $d\kappa$ term in the first transformation.

The symmetry (\ref{symm}, \ref{symm2}) is reducible and it is
convenient to exhibit the original transformation and the subsequent
ones as part of a long exact sequence on spaces of horizontal forms
$\Omega^{q}_{H} \oplus 
\Omega^{q-1}_{H} \rightarrow \Omega^{q+1}_{H} \oplus
\Omega^{q}_{H} $
\be
\Lambda^{(0)} \stackrel{d_\mathcal{A}}{\longrightarrow}
(\Lambda^{(1)}, \, \Lambda_{(0)})
\stackrel{d_\mathcal{A}}{\longrightarrow} (\Lambda^{(2)}, \,
\Lambda_{(1)})\stackrel{d_\mathcal{A}}{\longrightarrow} \dots
\stackrel{d_\mathcal{A}}{\longrightarrow} (\Lambda^{(p-1)}, \,
\Lambda_{(p-2)})\stackrel{d_\mathcal{A}}{\longrightarrow}(\Phi^{(p)}, \,
\Phi_{(p-1)})\label{sequence}
\ee
We now remind the reader of why the 
complete Batalin-Vilkovisky approach is required in order to take care of these reducible 
symmetries when using covariant gauges, as this will allow us to
explain why the algebraic gauge conditions that we choose allow for a
significantly reduced set of fields, compared with the complete set given by the appropriate
Batalin-Vilkovisky triangles.

Recall that if you have a $p$-form $B$ and $p-1$
form ghost $\omega$, anti-ghost $\overline{\omega}$ and multiplier field $\pi$
their transformations are, in the first instance,
\be
QB= d_{\mathcal{A}}\omega, \quad Q\omega= 0 , \quad Q\overline{\omega} = \pi, \quad
Q\pi =0 \label{brst}
\ee
and the obvious covariant gauge fixing is
\be
Q\langle d_{\mathcal{A}}\overline{\omega}, \, * B\rangle = \langle
d_{\mathcal{A}}\pi , \, * B\rangle 
- \langle d_{\mathcal{A}}\overline{\omega} , \, * d_{\mathcal{A}}\,
\omega \rangle \label{gf}
\ee
for some metric which provides the Hodge operator $*$ on the
manifold. The need for the complete triangle can be traced to the fact that
$d_{\mathcal{A}}\omega$,  $d_{\mathcal{A}}\overline{\omega}$ and
$d_{\mathcal{A}}\pi$ have degeneracy given by shifting the fields by
covariantly exact pieces. These symmetries arise for the anti-ghost and
multiplier fields because of the gauge fixing conditions themselves (\ref{gf}), while
for the ghosts they are an inherent part of the original
transformations (\ref{brst}). A prudent choice of non-covariant gauge
fixing conditions may mean that there 
will be no extra symmetries for the anti-ghosts and multiplier
fields which in turn means that one may be able to ignore the  plethora of ghosts for ghosts.

At this point it is convenient to regard the horizontal fields as being
split according to the orthogonal decomposition (\ref{par-perp})
\be
\Omega_{H}^{*}(M,E) = \Omega_{\perp}^{*}(M,E) \oplus \Omega_{\parallel}^{*}(M,E)
\ee
where $\Omega_{\parallel}^{*}(M,E)$ is the kernel of $D_{\phi}$ acting
on $\Omega_{H}^{*}(M,E) $. This split is implicitly understood with
$\Psi \in \Omega^{p}_{H}(M,E)$
\be
\Psi = \Psi_{\perp} + \Psi_{\parallel}, \quad \Psi_{\perp} \in
\Omega_{\perp}^{p}(M,E),\quad \Psi_{\parallel} \in
\Omega_{\parallel}^{p}(M,E)
\ee
(there is no difference between superscripts and subscripts of $\perp$
and $\parallel$ and the positions are chosen for legibility).

The form of the symmetries suggests we start with the fields at the
left of (\ref{sequence}) and work our way to the right.  The first
transformation is
\be
\delta \Lambda^{(1)} = d_{A}^{H}\Lambda^{(0)},
\quad \delta \Lambda_{(0)} = D_{\phi}\Lambda^{(0)} 
\ee
Using $\Lambda^{(0)}_{\perp}$ allows us to set $
\Lambda_{(0)}^{\perp}=0$. $\Lambda^{(0)}_{\parallel}$ is  still
available and only enters in the variation of the parallel
components of $\Lambda^{(1)}$ (we may then use a Landau type gauge for
these components). We are then left with $\Lambda^{(1)}_{\perp}$, $\Lambda^{(1)}_{\parallel}$ and
$ \Lambda_{(0)}^{\parallel}$ . These are the available
symmetries at the next step to the right of (\ref{sequence}). It is
important to note that thus far the gauge fixing is on $ \Lambda_{(0)}$.

On going one step to the right in (\ref{sequence}) we have
\be
\delta \Lambda^{(2)} = d_{A}^{H}\Lambda^{(1)} + d\kappa \, \Lambda_{(0)},
\quad \delta \Lambda_{(1)} = D_{\phi}\Lambda^{(1)} -
d_{A}^{H}\Lambda_{(0)} \label{level2}
\ee
Given that $ \Lambda_{(0)}\in
\Omega^{0}_{\parallel}(M, E)$ then, as was the case with $
\Lambda^{(0)}$, one may use
$\Lambda^{(1)}_{\perp}$ to impose the
condition that $\Lambda_{(1)}\in
\Omega^{1}_{\parallel}(M, E)$ (i.e. $\Lambda_{(1)}^{\perp}=0$). After
this gauge fixing the only free modes for 
$\Lambda^{(1)}$ are its parallel components. The left over
transformations in (\ref{level2}) are all in $\Omega^{*}_{\parallel}(M, E)$ and all of the gauge
choices have, thus far, been on $\Lambda_{(1)}$.

This repeats at each stage so that the gauge conditions are on
$\Lambda_{(q)}$ while the symmetries that are left are in $\Omega^{*}_{\parallel}(M, E)$
and of the form
\be
\delta \Lambda^{(q)} = d^{H}_{A}\Lambda^{(q-1)} + d\kappa \, \Lambda_{(q-2)},
\quad \delta \Lambda_{(q-1)} = 
- d^{H}_{A}\Lambda_{(q-2)} \label{levelq}
\ee
with all fields in $\Omega^{*}_{\parallel}(M, E)$.

These conditions can now be imposed on the fields in $\Omega^{*}_{\perp}(M,
E)$ without the full might of the Batalin Vilkovisky
procedure as they are fully algebraic. 
The typical gauge fixing term takes the form
\begin{align}
\sum_{q=0}^{p-1}Q\left(\langle \overline{\omega}^{(q)}_{\perp}, \, *
  \omega_{(q)} ^{\perp}\rangle \right) & =
\sum_{q=0}^{p-1}\langle \pi^{(q)}_{\perp}, \, * \omega_{(q)}
                                         ^{\perp}\rangle  - \sum_{q=0}^{p-1}
\langle \overline{\omega}^{(q)}_{\perp}, \, * \left( D_{\phi}
                                         \omega^{(q)}_{\perp}
                                         -d_{A}^{H}\omega_{(q-1)}^{\perp}\right)
                                         \rangle 
\nonumber  \\
  & \simeq \sum_{q=0}^{p-1} \langle \pi^{(q)}_{\perp}, \, *
    \omega_{(q)} ^{\perp}\rangle - \sum_{q=0}^{p-1}
\langle \overline{\omega}^{(q)}_{\perp}, \, * D_{\phi} \omega^{(q)}
    _{\perp}\rangle \label{gf-ghost}
\end{align}
where $\omega_{(p-1)}^{\perp}=B_{(p-1)}^{\perp}$, $\omega_{(-1)}^{\perp}=0$ and in the second line we
have used the delta function constraints imposed by the multiplier
fields that the $\omega_{(q)}^{\perp}=0$. There are no differential operators that
act on the multiplier fields 
and we have chosen $D_{\phi}$ to have no zero modes on the perpendicular spaces so that neither
$\pi_{\perp}$ nor $\overline{\omega}_{\perp}$ have any extra symmetry here. 
Consequently, no extra ghosts for ghosts (and their associated fields) are required
to compensate such redundant symmetries.

Of course one could have also used the complete Batalin-Vilkovisky
algorithm and the complete set of fields in the Batalin-Vilkovisky
triangle to arrive at the same result in the end. Here is an example
to illustrate this.

\begin{example}\label{example}{\rm{We show here how the complete Batalin-Vilkovisky
      triangle leads to (\ref{gf-ghost}) for a 2-form $B$ for the part
      in $\Omega^{*}_{\perp}(M, E)$. The ghost system
      is
      \be
      \begin{array}{llcrr}
        &  & B &  &\\
      &  \overline{\omega}& & \omega &\\
      \beta' & & \overline{\beta} & & \beta
      \end{array} \longrightarrow
      \begin{array}{rrcll}
      &  & B^{(2)}+ \kappa B_{(1)} &  &\\
      &  \overline{\omega}^{(1)}+ \kappa \overline{\omega}_{(0)}& &
                                                                    \omega^{(1)}+ \kappa \omega_{(0)} &\\ 
      \omega'^{(0)} & & \overline{\omega}^{(0)} & & \omega^{(0)}  
      \end{array} \label{ghost-triangle}
      \ee
      The nilpotent BRST transformations are then
\be
      \begin{aligned}
        & QB^{(2)}  = d_{A}^{H}\omega^{(1)} + d\kappa \, \omega_{(0)} , \quad Q
                   B_{(1)} = D_{\phi}\omega^{(1)} - d_{A}^{H} \omega_{(0)} \\
                   & Q\omega^{(1)}  = d_{A}^{H} \, \omega^{(0)}, \quad Q \omega_{(0)}=
                              D_{\phi} \, \omega^{(0)}, \quad Q \omega^{(0)} =0
                              \\
        & Q\overline{\omega}^{(1)} = \pi^{(1)}, \quad Q\overline{\omega}_{(0)} =
                             \pi_{(0)}, \quad Q\pi^{(1)}=0, \quad Q\pi_{(0)}=0 \\
        & Q \omega'^{(0)} = \pi'^{(0)}, \quad Q \pi'^{(0)} =0, \quad Q
          \overline{\omega}^{(0)} = 
              \pi^{(0)}, \quad Q\pi^{(0)} =0.
      \end{aligned}\
\ee
 The fields that are not present in (\ref{gf-ghost}) are $(\omega'^{(0)} , \,
 \pi'^{(0)}, \overline{\omega}_{(0)}, \pi_{(0)})$ so one needs to append to
 (\ref{gf-ghost}) terms involving this set. However, we have already
 achieved the gauge fixing that is available to us (in $\Omega^{*}_{\perp}(M, E)$) and none
 of the quadruplet transform with derivatives so one may simply add
 the ghost number neutral term
 \be
 Q\langle \omega'^{(0)}_{\perp}, \, *
 \overline{\omega}_{(0)}^{\perp}\rangle =  \langle \pi'^{(0)}_{\perp},
 \, *
 \overline{\omega}_{(0)}^{\perp}\rangle - \langle
 \omega'^{(0)}_{\perp}, \, *\pi_{(0)}^{\perp} \rangle 
 \ee
 which shows that one is allowed to ignore these fields.
    }}
\end{example}

The path integral should not depend on the metric that is chosen for
the gauge fixing conditions, since it  always enters in a BRST exact
term, so that one may use this freedom to pick a metric on $M$ adapted
to the needs of the problem. The metric of choice here on $M$ is
$g_{M}$ as given in (\ref{gM}). In particular between horizontal
fields the Hodge star operator is $*_{H}$ (\ref{hodgeH}).

There are also symmetries that do not act, corresponding to zero modes
that is to elements of $\mathrm{H}^{*}_{\mathcal{A}}(M,E)$. For
example, the fields $\Lambda_{(q-2)}^{\parallel}$ and $\Lambda^{(q-1)}_{\parallel}$ with $d^{H}_{A}\Lambda_{(q-2)}^{\parallel} =0$
and  $d\kappa
\Lambda_{(q-2)}^{\parallel}= d_{A}^{H}\Lambda^{(q-1)}_{\parallel}$ correspond to
zero modes of the covariant derivative and they do not appear in
(\ref{levelq}). The treatment of such modes using a BRST analysis can
be found in \cite{BT-BF, BKT} and we do not repeat that here.

The symmetries that remain unaccounted for, those that appear in 
(\ref{levelq}), are all in
$\Omega^{*}_{\parallel}(M, E)$ and we should employ the complete
Batalin-Vilkovisky triangle for these when gauge fixing them. 

\begin{example}\label{example-zm}{\rm{We
continue Example \ref{example} to include the parallel sector in order to exhibit the relevant
features. The transformations are (we drop the parallel sub- and super-scripts)
\be
\begin{aligned}
        & QB^{(2)} = d_{A}^{H}\omega^{(1)} +
          d\kappa \, \omega_{(0)}  , \quad Q 
                   B_{(1)}=  - d_{A}^{H} \omega_{(0)} \\
                   & Q\omega^{(1)} = d_{A}^{H} \,
                     \omega^{(0)}, \quad Q
                     \omega_{(0)}=0, \quad Q \omega^{(0)}=0
                              \\
        & Q\overline{\omega}^{(1)}=
          \pi^{(1)}, \quad
          Q\overline{\omega}_{(0)}= 
                             \pi_{(0)}, \quad
          Q\pi^{(1)}=0, \quad Q\pi_{(0)}=0
          \\ 
        & Q \omega'^{(0)}= \pi'^{(0)}, \quad Q \pi'^{(0)}=0, \quad Q
          \overline{\omega}^{(0)} = 
              \pi^{(0)}, \quad Q\pi^{(0)}=0.
 \end{aligned}
\ee
and in order to gauge fix the non-harmonic modes we form the action
 \begin{align}
 \int_{M}\kappa   \; &  Q\left( \langle\overline{\omega}^{(1)}, \, 
   d_{A}^{H}*_{H}B^{(2)} \rangle + \langle \overline{\omega}_{(0)}, \, 
   d_{A}^{H}*_{H}B_{(1)} \rangle \right.\nonumber\\
   & \left. +
   \langle \overline{\omega}^{(0)}, \, d_{A}^{H}*_{H}\omega^{(1)}
     \rangle +
   \langle \omega'^{(0)}, \,
     d_{A}^{H}*_{H}\overline{\omega}^{(1)}\rangle \right)
 \end{align}
 The gauge fixing terms are
 \begin{align}
 \int_{M}\kappa   \; &  \left( \langle\pi^{(1)}, \, 
   d_{A}^{H}*_{H}B^{(2)} \rangle + \langle \pi_{(0)}, \, 
                       d_{A}^{H}*_{H}B_{(1)} \rangle +
                    \right.\nonumber\\
  & \left. 
   \langle \pi^{(0)}, \, d_{A}^{H}*_{H}\omega^{(1)}
     \rangle +
   \langle \pi'^{(0)}, \,
     d_{A}^{H}*_{H}\overline{\omega}^{(1)}\rangle  - \langle \omega'^{(0)}, \,
     d_{A}^{H}*_{H}\pi^{(1)}\rangle\right)
 \end{align}
 while the `ghost' terms are
 \begin{align}
& -\int_{M}\kappa   \; \left( \langle \overline{\omega}^{(1)}, \, 
   d_{A}^{H}*_{H} (d_{A}^{H} \omega^{(1)} + d\kappa \, \omega_{(0)} )\rangle
   +  \langle \overline{\omega}_{(0)}, \, 
   d_{A}^{H}*_{H} d_{A}^{H}\omega_{(0)} \rangle - \langle \overline{\omega}^{(0)}, \,
     d_{A}^{H}*_{H}d_{A}^{H}\omega^{(0)}\rangle \right)
 \end{align}
 
 There is only one mixing term between the ghosts of the two BV
 triangles associated with $B^{(2)}$ and $B_{(1)}$ which in combined
 form appear in (\ref{ghost-triangle}), and it appears in the combination 
 \be
 -\int_{M}\kappa   \; \langle \overline{\omega}^{(1)}, \, 
   d_{A}^{H}*_{H} d\kappa \, \omega_{(0)} )\rangle \label{mixing}
   \ee
Note that as the gauge fixing condition on $\omega^{(1)}$ and the
integral over $\overline{\omega}^{(1)}$ imply that $\omega^{(1)}$ is
harmonic the mixing term (\ref{mixing}) vanishes
regardless of the properties of $\omega_{(0)} $ which
 means that one could just as well set the mixing term
 $\langle \overline{\omega}^{(1)}, \, d_{A}^{H}*_{H} d\kappa \,
 \omega_{(0)}\rangle $ to
 zero from the outset.  All in all this means that we could have set the $d\kappa$
 term in the gauge symmetries to zero to be left with two independent
 and standard sets of gauge symmetries for $B^{(2)}$ and $B_{(1)}$
 respectively at this level. But now we must address the harmonic
 modes as every field in this sector potentially
 has harmonic modes and in particular we must determine which of those
 modes is part of the symmetry group and which is actually a zero
 mode. This questions devolves in this case to understanding the
 harmonic modes of $\omega_{(0)} $ as they are the ones that will
 appear in the gauge algebra. Let $\widehat{d\kappa}$ be the harmonic, with respect to $\Delta_{H}$,
 part of $d\kappa$. If $\omega_{(0)} $ is harmonic then so
 too is $\widehat{d\kappa} \, \omega_{(0)} $ as it is obviously closed with
 respect to $d_{A}^{H}$ while $d_{A}^{H}*_{H} (\widehat{d\kappa} \,
 \omega_{(0)} )= (d_{H}*\widehat{d\kappa})\, \omega_{(0)} =0$. As $\widehat{d\kappa}\,
 \omega_{(0)} $ is harmonic it is part of the gauge symmetry and we
 can use it to partially fix some of the harmonic modes of
 $B^{(2)}$. To that end add to the action
 \be
 Q\int_{M}\kappa \, \langle \widehat{d\kappa} \, \overline{\omega}_{(0)}, *_{H}
 B^{(2)}\rangle = \int_{M}\kappa \, \left(\langle \widehat{d\kappa} \, \pi_{(0)}, *_{H}
 B^{(2)}\rangle - \widehat{d\kappa} *_{H} \widehat{d\kappa}\, \langle \overline{\omega}_{(0)}, 
 \omega_{(0)}\rangle \right) \label{gf-harm}
\ee
where all the fields in (\ref{gf-harm}) are harmonic. Note that this
gauge fixing term is now the only one that mixes the BV triangles. All the other
harmonic modes of $B^{(2)}$ and of the other fields are zero modes and
may dealt with as such \cite{Ray-Singer-Analytic, BKT}.
}}
\end{example}

 Turning to the general situation for the parallel fields,
 we can split off the BV triangles
 for the field $B^{(p)}$ and that of $B_{(p-1)}$, even though they are
 not independent due to the mixing terms. All the covariant gauge fixing is, as
 above,  with $d_{A}^{H}*_{H}$ acting on the fields. 
 However, integrating out the
 anti-fields and multiplier fields of the $B_{(p-1)}$ triangle (apart
 from generating determinants of the the $\Delta_{A}^{H}$ Laplacian)
 sets all the ghost fields to be harmonic- including the ones that would make
 an appearance in the symmetries of the $B^{(p)}$ triangle. Up to this
 point one may ignore the $d\kappa \Lambda_{(r)}$ terms completely and the
 triangles are essentially independent. This is exactly what we found
 in Section \ref{section4} by the scaling argument. However, in order
 to treat the harmonic modes one must first ascertain if they are part of
 the gauge symmetry or not. If they are part of the gauge symmetry then a gauge fixing of the
 type (\ref{gf-harm}) is to be used while if they are not part of the
 gauge symmetry then they are zero modes and one may follow
 \cite{Ray-Singer-Analytic, BKT} in order to deal with them.

\rnc{\Large}{\normalsize}

\end{document}